\documentclass[
    aps,
    physrev,
    twocolumn,
    10pt
]{revtex4-2}

\usepackage{array}[=2016-10-06] 

\usepackage{graphicx}
\usepackage{amsmath}
\usepackage{natbib} 
\usepackage{amssymb}
\usepackage{color}

\usepackage{hyperref}

\usepackage{enumitem}  

\usepackage[nolist]{acronym}

\usepackage{physics}

\AtBeginDocument{\RenewCommandCopy\qty\SI}

\usepackage{siunitx}  
\DeclareSIUnit\parsec{pc}
\DeclareSIUnit{\Mpc}{\mega\parsec}

\newcommand{\Msun}{\,\mathrm{M}_{\odot}}

\graphicspath{{./figs/}}

\usepackage[capitalise,noabbrev]{cleveref}

\newcommand{\mismatch}{\mathfrak{M}}
\newcommand{\chieff}{\chi_{\rm eff}}
\newcommand{\snr}{\rho}
\newcommand{\params}{\theta}

\newcommand{\normal}{\mathcal{N}}
\newcommand{\like}{\mathcal{L}}  

\emergencystretch=\maxdimen
\begin{document}


\title{Gravitational-wave parameter estimation with machine-learning generated surrogate waveforms}

\author{Suyog Garg}
\email{gargsuyog@g.ecc.u-tokyo.ac.jp}
\affiliation{Department of Physics, The University of Tokyo, Bunkyo-ku, Tokyo 113-0033 JAPAN}
\affiliation{Research Center for Early Universe, The University of Tokyo, Bunkyo-ku, Tokyo 113-0033 JAPAN}

\author{Kipp Cannon}
\affiliation{Department of Physics, The University of Tokyo, Bunkyo-ku, Tokyo 113-0033 JAPAN}
\affiliation{Research Center for Early Universe, The University of Tokyo, Bunkyo-ku, Tokyo 113-0033 JAPAN}


\date{\today}

\begin{abstract}
    The worldwide network of gravitational-wave detectors have detected more than 350 binary coalescence events till date. Future third-generation detectors, like Einstein telescope, are expected to detect orders-of-magnitude more signals from sources with more complicated characteristics, including eccentric orbits and high-mass ratio binaries. 
    It is well-established that the computational cost of parameter estimation for signals from these kinds of sources will be extremely high. In particular, the process could be sped-up if generating theoretical waveform predictions, used for likelihood calculation becomes faster. Recently, various machine-learning techniques has been proposed to this end.
    In this work, we propose a two-stage deterministic conditional-autoencoder model for generating four-parameter SEOBNRv4 waveforms. The first-stage of the model generates amplitude and phase series of the waveform, while the second-stage calibrates the residual error in the predictions. Our model achieves a median mismatch of around $10^{-2}$ with the target polarization waveforms, while the calibrated amplitude/phase series achieve $10^{-6}$ level cosine distance error.
    We then propose a waveform conditioning step to enable use of these surrogate waveforms for downstream parameter estimation tasks. Finally, we perform extensive parameter estimation tests, with ML and EOB waveform injections and try to recover posterior estimates for the source parameters. We find that when ML waveforms are used to recover EOB target parameter estimates, the inferred posterior have some systematic bias. This inherent bias can be estimated and corrected for, and then importance reweighting of posterior samples can enable use of low-accuracy surrogate waveforms at low SNRs.
\end{abstract}


\maketitle

\acrodef{utokyo}[UTokyo]{the University of Tokyo}
\acrodef{lvkc}[LVK collaboration]{LIGO--Virgo--KAGRA collaboration}
\acrodef{kld}[KLD]{Kullback--Leibler divergence}

\acrodef{resceu}[RESCEU]{Research Center for Early Universe}
\acrodef{kias}[KIAS]{Korea Institute of Advanced Studies}
\acrodef{mit}[MIT]{Massachusetts Institute of Technology}
\acrodef{aei}[AEI]{Max Planck Institute for Gravitational Physics (Albert Einstein Institute)}
\acrodef{ego}[EGO]{European Gravitational Observatory}
\acrodef{icrr}[ICRR]{Institute for Cosmic Ray Research, University of Tokyo}
\acrodef{naoj}[NAOJ]{National Astronomical Observatory of Japan}
\acrodef{ifae}[IFAE]{Institut de Física d'Altes Energies}
\acrodef{igwn}[IGWN]{International Gravitational Wave Network}
\acrodef{cern}[CERN]{European Organization for Nuclear Research}
\acrodef{gwosc}[GWOSC]{Gravitational Wave Open Science Center}
\acrodef{gwtc}[GWTC]{Gravitational-Wave Transient Catalog}
\acrodef{cpu}[CPU]{central processing unit}
\acrodef{gpu}[GPU]{graphics processing unit}
\acrodef{hpc}[HPC]{high performance computing}

\acrodef{gw}[GW]{gravitational-wave}
\acrodef{ligo}[LIGO]{Laser Interferometer Gravitational-Wave Observatory}
\acrodef{virgo}[Virgo]{Virgo Interferometer}
\acrodef{kagra}[KAGRA]{Kamioka Gravitational Wave Detector}
\acrodef{lisa}[LISA]{Laser Interferometer Space Antenna}
\acrodef{et}[ET]{Einstein Telescope}
\acrodef{ce}[CE]{Cosmic Explorer}
\acrodef{ligoindia}[LIGO-India]{Laser Interferometer Gravitational-Wave Observatory in India}
\acrodef{lvk}[LVK]{LIGO--Virgo--KAGRA}
\acrodef{cbc}[CBC]{compact binary coalescence}
\acrodef{bbh}[BBH]{binary black hole}
\acrodef{bns}[BNS]{binary neutron star}
\acrodef{nsbh}[NSBH]{neutron star--black hole}
\acrodef{pe}[PE]{parameter estimation}
\acrodef{snr}[SNR]{signal-to-noise ratio}
\acrodef{psd}[PSD]{power spectral density}
\acrodef{asd}[ASD]{amplitude spectral density}
\acrodef{imr}[IMR]{inspiral-merger-ringdown}
\acrodef{qnm}[QNM]{quasi-normal mode}

\acrodef{ml}[ML]{machine-learning}
\acrodef{dl}[DL]{deep learning}
\acrodef{ae}[AE]{autoencoder}
\acrodef{vae}[VAE]{variational autoencoder}
\acrodef{cae}[CAE]{conditional autoencoder}
\acrodef{cvae}[CVAE]{conditional variational autoencoder}
\acrodef{ann}[ANN]{artificial neural network}
\acrodef{cnn}[CNN]{convolutional neural network}
\acrodef{rnn}[RNN]{recurrent neural network}
\acrodef{mse}[MSE]{mean squared error}
\acrodef{kl}[KL]{Kullback--Leibler}
\acrodef{sbi}[SBI]{simulation-based inference}
\acrodef{npe}[NPE]{neural posterior estimation}
\acrodef{nde}[NDE]{neural density estimation}
\acrodef{nf}[NF]{normalizing flow}
\acrodef{pinn}[PINN]{physics-informed neural network}

\acrodef{mcmc}[MCMC]{Markov chain Monte Carlo}
\acrodef{ns}[NS]{nested sampling}
\acrodef{lh}[LH]{likelihood function}
\acrodef{map}[MAP]{maximum a posteriori}
\acrodef{mle}[MLE]{maximum likelihood estimation}
\acrodef{pdf}[PDF]{probability density function}
\acrodef{cdf}[CDF]{cumulative distribution function}
\acrodef{pp}[PP]{probability-probability}
\acrodef{ci}[CI]{credible interval}
\acrodef{hpd}[HPD]{highest posterior density}
\acrodef{js}[JS]{Jensen--Shannon}

\acrodef{gr}[GR]{general relativity}
\acrodef{nr}[NR]{numerical relativity}
\acrodef{pn}[PN]{post-Newtonian}
\acrodef{eob}[EOB]{effective one-body}
\acrodef{seobnr}[SEOBNR]{effective one-body numerical relativity waveform family (spin-aligned/precessing)}
\acrodef{imrphenom}[IMRPhenom]{phenomenological inspiral-merger-ringdown waveform family}



\section{Introduction}
\label{sec:intro}

The direct detection of \ac{gw} in 2015 by the \ac{lvk} collaboration ushered in the era of \ac{gw} astronomy \cite{abbott2016}. Since then, the number of successfully detected \ac{gw} events has reached more than 350 \ac{cbc} detections, in the five \acp{gwtc} released till date \cite{abbott2019,collaboration2021,collaboration2023,theligoscientificcollaboration2025a,LVK2026GWTC5Catalog}. In the near-future this number is expected to further increase, with the commissioning of third-generation ground-based detectors like the Einstein telescope \cite{abac2025b} and Cosmic Explorer \cite{hall2021}, and space-borne detectors, LISA \cite{colpi2024}, Taiji \cite{luo2020a,luo2021} and TianQin \cite{luo2016,liLi2025a}.

The detectable \ac{gw} signals for these third-generation detectors, and even in future observation runs of the current \ac{gw} detector network, will originate from sources with more complicated orbital characteristics, for instance, eccentric and precessing system \cite{xuan2023,gupta2024}. It is well-established that probing these orbital characteristics from the \ac{gw} signal morphology, can shed light on the population-level distribution of the binaries and lead to other astrophysical implications \cite{collaboration2025}. Now, full \acl{pe} of signals with relatively complicated source characteristics and signal morphology, than compared to, say, a dominant-mode quasi-circular \ac{bbh} waveform, is computationally expensive. One leading cause for this is the waveform generation time \cite{garg2026a}.

There can be two basic ways to speed-up the \acl{pe} process, either having faster waveform generation models for likelihood calculation, or direct estimation of parameter posterior, without having to calculate the likelihood explicitly. More efficient sampling algorithms \cite{tiwari2023}, relative-binning with a reference waveform on a coarser frequency grid \cite{zackay2018,dai2018,leslie2021,krishna2023,narola2023,narola2024}, and other techniques \cite{vinciguerra2017,xia2021,dax2021,gabbard2022,dax2023,nitz2025,peng2025}, can then be combined with these to have an overall speed-up in the \acl{pe} process. In particular, faster waveform generation methods are not only helpful for \acl{pe}, but also for other applications, like tests of gravity, template-bank generation and other theoretical studies.

Recently there has been a significant progress on development of \ac{ml} methods for gravitational waveform modelling \cite{cuoco2020,cuoco2025,Zhao2025AIReview}. These include surrogate waveform models of limited accuracy, to high-accuracy surrogates for some specific waveform approximants \cite{blackman2017,liao2021,lee2021,khan2021,chua2021,schmidt2021,thomas2022,shi2024,grimbergen2024,chatterjee2024,gadre2024,sun2025a,garg2025a,freitas2025,garg2026a,whittall2026,theodoropoulos2026}. People have also tried \acl{pe} with these surrogate models by inferencing source parameter for few significant events from the catalog \cite{tissino2023,cuoco2025,garg2025a,whittall2026}. A comprehensive \acl{pe} analysis with \acl{ml} surrogate waveforms, characterizing any inherent inference bias in these surrogate waveforms, is currently lacking. In this work we try to accomplish this and show that just the accuracy criteria of surrogate waveform output alone is not sufficient to imply its use for downstream \acl{pe} tasks.

\citet{garg2026a} introduced a \ac{cvae} model for fast generation of \ac{eob} SEOBNRv4 \cite{BuonannoDamour1999,damour2009,nagar2018} gravitational waveforms from aligned-spin \acl{bbh} sources. This model achieves batched-\ac{gpu} waveform generation speed of $\order{10^{4}}$ waveforms in $\order{10^{-1}}$\,\unit{\second}, however, the generated waveforms come at the cost of a lowered accuracy at $\order{10^{-2}}$ mismatch. This model is also stochastic in nature due to the latent space been a learned normal distribution, and thus has an additional $\order{10^{-2}}$ uncertainty in the reconstructed waveforms.

In this work, we introduce an alternative set-up with a \ac{cae} \cite{kingma2019,kingma2022} waveform generation model which removes the uncertainty in the generated outputs arising due to the stochastic nature of a variational model.
We further perform a calibration of the \acl{ml} generated outputs by predicting the residual error with a second-stage calibrator model. This calibration stage improves the accuracy of the \acl{ml} model outputs by around two orders-of-magnitude, although the accuracy of the reconstructed waveform polarizations remain on-par or only slightly better than \citet{garg2026a}.
We then posit waveform conditioning as a necessary intermediate step before downstream use of surrogate waveforms. Finally, we perform extensive \acl{pe} test with the conditioned waveform outputs of the two-stage waveform generation and calibration model.

Results presented in this work demonstrate that even low-accuracy \acl{ml} model based surrogate waveforms can be used for \acl{gw} \acl{pe} applications, although the estimated posteriors tend to have some inherent inherence bias. We propose that this bias can be corrected, following which, an importance-reweighting of the posterior samples can lead to much better posterior estimates \cite{veach1997,elvira2019}.


\section{Data preprocessing}

The parameter ranges for our priors are set as,
\begin{itemize}[nosep]
    \item Mass: $m_1,m_2\in[5,75]\,\Msun$, with $m_2/m_1<10\,\Msun$
    \item Aligned-spins: $\chi_1(z),\chi_2(z)\in[-0.99,0.99]$.
\end{itemize}
This chosen mass range is a simple approximation based on the O1--O2 \ac{gwtc} mass ranges for the \aclp{bbh} \cite{abbott2019,collaboration2021}. Results and conclusions presented in our analysis are expected to remain valid even for different prior ranges and training data.
For each set of parameters from the priors, we obtain our target waveforms for the \ac{eob} approximant \texttt{SEOBNRv4} from the \texttt{LALSuite} library \cite{LALSuite2018} via \texttt{PyCBC} \cite{alexnitz2024} and \texttt{SwigLAL} \cite{wette2020} interfaces. These waveforms serve as the target data for training, validation and testing of outputs generated by our \acl{ml} waveform model.

Similar to \citet{garg2026a}, all the waveforms in our dataset are sampled at a sampling rate of \qty{8192}{\hertz} with a fixed-duration of \qty{1}{\second}. 
However, the duration of the \acl{gw} signal from sources with different parameters is not constant at a fixed lower frequency cut-off ($f_{\rm low}$). 
Specifically, time to coalescence is proportional to the binary mass, $t_c\propto f_{\rm low}^{-8/3}/\mathcal{M}^{-5/3}$, where $\mathcal{M}=(m_1m_2)^{3/5}/(m_1+m_2)^{1/5}$ is the chirp mass.
Thus, to obtain fixed-duration signals for different source parameters, we apply a dynamic lower frequency cut-off, similar to \citet{garg2026a}. The final waveforms used in our analysis have a $f_{\rm low}$ range of \qtyrange{11}{60}{\hertz}, dynamically chosen based on the mass of the binary.

In this work, we only consider the dominant $h_{22}$ mode of the gravitational waveform, and decompose the real-valued polarization time-series $h_{+}(t)$ and $h_{\times}(t)$ into amplitude and phase as follows,
\begin{align}
    A(t) &= \sqrt{h_{+}^2 + h_{\times}^2} \nonumber \\
    \phi(t) &= \mathrm{unwrap}\, [ \arctan (h_{+}, h_{\times}) ].
\end{align}%
These amplitude and phase series are the inputs and the targets for our \acl{ml} waveform model. We further normalize the inputs as, $x_{\rm norm}=(x-\mu)/(\sigma+\epsilon)$, where $\mu$ and $\sigma$ are respectively the mean and the standard deviation, and $\epsilon$ is a small random number to compensate for divide-by-zero errors. The target amplitude and phase series are kept unnormalized.


\section{Methodology}

Our \acl{ml} waveform model consists of two stages: 1) waveform generation stage, and 2) residual calibration stage. 
This two-stage \texttt{FlexCAEPhase} framework separates the problem of fast waveform generation from the problem of correcting systematic model errors. In practice, this improves the reconstructed waveform in the regions where the first-stage approximation is less accurate, especially near the late inspiral and merger where small phase and amplitude errors can have a larger effect on the waveform mismatch. 

Both the waveform generation model and the calibration model are trained for 100 epochs on a Nvidia A100 80\unit{\giga\byte} \ac{gpu} using \texttt{PyTorch} \cite{paszke2019} and \texttt{CUDA} \cite{nickolls2008}.
Both stages use the \texttt{ADAM} \cite{kingma2017} optimization with learning-rate scheduling, and the final calibration model is selected according to a low validation loss value. To reduce computational cost, the calibration inputs and residual targets are precomputed and stored in HDF files \cite{thehdfgroup2026}, so that the second-stage model trains directly from disk without repeatedly evaluating the waveform generator.

\subsection{Waveform generation model}

In the first stage, the waveform generation model takes in the source parameters as the input and outputs a two-channel $[A(t),\phi(t)]$ series. Contrary to \citet{garg2026a}, we now do not work with a variational model, in which the learned latent space is distribution. Instead, our waveform generation model maps the inputs to a latent space consisting of discrete latent vectors. This requirement removes the stochasticity in the model generated outputs, as was observed in \citet{garg2026a}.

For the aligned-spin configuration considered in this work, the input parameter vector is written as $\params=[m_1,m_2,\chi_{1}(z),\chi_{2}(z)]$,
where $m_1$ and $m_2$ are the component masses and $\chi_{1}(z)$, $\chi_{2}(z)$ are the dimensionless spin components aligned with the orbital angular momentum. The model predicts an intermediate waveform representation, namely the normalized amplitude and instantaneous phase,
\begin{equation}
    \widehat{\mathbf{y}}_{\rm ML}
    = [\widehat{A}_{\rm ML}(t),\widehat{\phi}_{\rm ML}(t)].
\end{equation}
This representation is then converted into the two \acl{gw} polarizations, 
$\widehat{h}^{\rm ML}_{+}(t), \widehat{h}^{\rm ML}_{\times}(t)$, through the usual trigonometric decomposition of the strain.

The specific model architecture for training the waveform generation model, otherwise, remains the same as in \citet{garg2026a}. During training, the model consists of 2 encoders, 2 conditional-prior encoders and 1 decoder. One encoder encodes the target waveforms into latent representations, while the other encodes the normalization parameters. The two conditional-prior encoders likewise encode the source parameter labels $[m_1,m_2,\chi_{1}(z),\chi_2(z)]$, into latent vectors corresponding to each of the main encoders. The decoder then takes these discrete encoded latent vectors and tries to decode back the target waveforms from them. The training process aims to minimize the sum of the reconstruction error, i.e. the mean squared difference between the generated output and the target, and the cosine difference between the pairs of encoded latent vectors.

Once the model has trained, it has in effect learned the mapping from the source parameters to some latent vectors which decode back to a close approximation of the original target waveforms. 
On passing, we note that this learned latent vector space could be interpreted to find correlations between distinct latent vectors and the mismatch between the corresponding waveform under identical noise conditions.
To use the trained model for waveform generation, we simply pass in the source parameter labels to the conditional-prior encoders and obtain reconstructed waveform outputs.


\begin{table*}
\caption{Details of the network architecture used in the waveform generation first-stage of the FlexCAEPhase model.}
\label{tab:flexcaephase_model}
\begin{tabular}{llp{0.5\linewidth}}
\hline
Component & layers &   Notes \\
\hline

XEncoder
& pre-FC layers
& None
\\

& CNN layers 
& 
Conv1D (in=2,out=32,k=6,d=1) + GELU + MaxPool (k=4),
\par
Conv1D (in=32,out=32,k=6,d=1) + GELU + MaxPool (k=4)
\\

& post-FC layers 
& 
Linear (16324,256) + GELU,
Linear (256,256) + GELU,
Linear (256,48) + GELU
\\

&
& 
    \textit{
        Encodes inputs shape $(2,8191)$ to latent dimension $d_{\rm X}=24$.
    }
\\

KeyEncoder
& pre-FC layers
& Linear (8,500) + GELU,
Linear (500,500) + GELU,
Linear (500,8) + Identity
\\

& CNN layers 
& None
\\

& post-FC layers 
& None
\\

& 
& \textit{
    Encodes keys shape $(2,2)$ to latent dimension $d_{\rm key}=4$.
    }
\\

XConditional
& pre-FC layers
& Linear (4,128) + GELU,
Linear (128,256) + GELU,
Linear (256,128) + GELU,
Linear (128,48) + GELU,
\\

& CNN layers 
& None
\\

& post-FC layers 
& None
\\

& 
& \textit{
    Conditional prior network for the waveform latent vector.
    }
\\

KeyConditional
& pre-FC layers
& Linear (4,128) + GELU
Linear (128,256) + GELU,
Linear (256,128) + GELU,
Linear (128,8) + GELU
\\

& CNN layers 
& None
\\

& post-FC layers 
& None
\\

& 
& \textit{
    Conditional prior network for the key latent vector.
    }
\\

Decoder
& pre-FC layers
& Linear (32,800) + Identity
\\

& CNN layers 
& Conv1D (in=1,out=64,k=5,d=3) + GELU + MaxPool (k=4),
\par
Conv1D (in=64,out=64,k=5,d=3) + GELU + MaxPool (k=4),
\par
Conv1D (in=64,out=64,k=5,d=3) + GELU + MaxPool (k=4)
\\

& post-FC layers 
& Linear (512,256) + GELU,
Linear (256,16382) + GELU
\\

&
& \textit{
    Decoder input uses concatenation of latent vectors of effective size 32.
    }
\\

\hline
\end{tabular}

\par\smallskip
\textit{Note:} The model uses GELU activations, batch size 1024, starting learning rate $5\times 10^{-4}$. The target representation is amplitude-phase, with normalized inputs and unnormalized targets. The total number of trainable parameters is $9,072,646$. The four conditioning labels correspond to the intrinsic binary parameters $[m_1,m_2,\chi_{1}(z),\chi_{2}(z)]$. The notation $d_{\rm X}$ and $d_{\rm key}$ denotes the latent dimensionality of the waveform and key representations, respectively.
For the layer architecture, `k' and `d' denote the kernel-size and the dilation respectively.
\end{table*}


\subsection{Residual calibration model}

\begin{table}[t]
\centering
\small
\setlength{\tabcolsep}{4pt}
\renewcommand{\arraystretch}{1.15}
\caption{Architecture of the residual calibration network used in the second-stage of the FlexCAEPhase model. The model takes a 6-channel input consisting of first-stage output amplitude/phase and the four normalized source parameters, and predicts two residual channels.}
\label{tab:calibration_cnn_arch}
\begin{tabular}{p{1.5cm}p{4.85cm}p{1.35cm}}
    \hline
    Layers & Configuration & Output shape \\

    \hline
    Input projection
    & Conv1D(in=6,out=128,k=7,p=3) & $128 \times N$ 
    \\

    Residual layers 
    & $5 \times$ 
    \par [Conv1D(in=128,out=128,k=7) 
    $\rightarrow$ GroupNorm(n=8) 
    \par
    $\rightarrow$ GELU 
    $\rightarrow$ Dropout(0) 
    \par
    $\rightarrow$ Conv1D(in=128,out=128,k=7) 
    \par
    $\rightarrow$ GroupNorm(n=8) 
    \par
    $\rightarrow$ skip-connection 
    $\rightarrow$ GELU
    ] 
    & $128 \times N$ 
    \\

    Output projection 
    & Conv1D(in=128,out=2,k=7,p=3);
    \par initialized to zero 
    & $2 \times N$ \\
    \hline
\end{tabular}
\par\smallskip
\footnotesize Total trainable parameters: $1{,}158{,}018$.
`p' denotes the padding size in the convolution layer.
\end{table}

Although the first-stage model provides a fast approximation to the target waveform family, small errors in amplitude and phase can lead to non-negligible mismatch in the final polarizations. To reduce these systematic errors, a second-stage residual calibration network is introduced. This calibration model is trained after the first-stage model has been fixed. The inputs to the second-stage model consist of the predicted \acl{ml} waveform and the corresponding physical parameters,
\begin{equation}
    \mathbf{x}_{\rm cal}^{(2)}
    = \left[
    \widehat{A}_{\rm ML}(t),
    \widehat{f}_{\rm ML}(t),
    m_1,
    m_2,
    \chi_{1}(z),
    \chi_{2}(z)
    \right],
\end{equation}
where the scalar parameters are repeated along the time dimension so that the complete input can be treated as a multi-channel one-dimensional sequence. The target for the second-stage model is the residual between the reference waveform representation and the first-stage prediction,
\begin{equation}
    \mathbf{r}^{(2)}(t)
    = \mathbf{y}_{\rm true}(t)
    - \widehat{\mathbf{y}}_{\rm ML}(t),
\end{equation}
with $\mathbf{y}=[A,f]$. The calibrated amplitude-phase representation is then obtained as
\begin{equation}
    \widehat{\mathbf{y}}_{\rm cal}(t)
    = \widehat{\mathbf{y}}_{\rm ML}(t)
    + \widehat{\mathbf{r}}^{(2)}(t),
\end{equation}
where $\widehat{\mathbf{r}}^{(2)}(t)$ is the residual predicted by the calibration network.

The calibration model is trained using a merger-amplitude weighted loss function, that penalizes errors at merger time high-amplitude region more heavily in comparison to elsewhere.
\begin{align}
    \mathcal{J}_{\rm mwMSE}
    &=
    \frac{\sum_{j=1}^{N} w_j \left(r_j-\hat{r}_j\right)^2}{\sum_{j=1}^{N} w_j},
    \nonumber \\
    w_j
    &=
    w_{\min}
    +
    (1-w_{\min})
    \left(
    \frac{|A^{\rm ML}_j|}{\max_k |A^{\rm ML}_k|+\epsilon}
    \right)^{\gamma},
\end{align}%
In this merger-weighted \ac{mse} loss function, the residual at each time sample is weighted by the normalized amplitude of the \acl{ml} waveform, so that points near the high-amplitude merger region receive larger penalties than points in the low-amplitude inspiral. A small floor value $w_{\min}$ is retained so that the early inspiral still contributes to the loss function, while the exponent $\gamma$ controls how sharply the weighting concentrates around the merger. 





\section{Results for waveform generation and calibration}

\begin{figure}
    \centering
    \includegraphics[width=\linewidth]{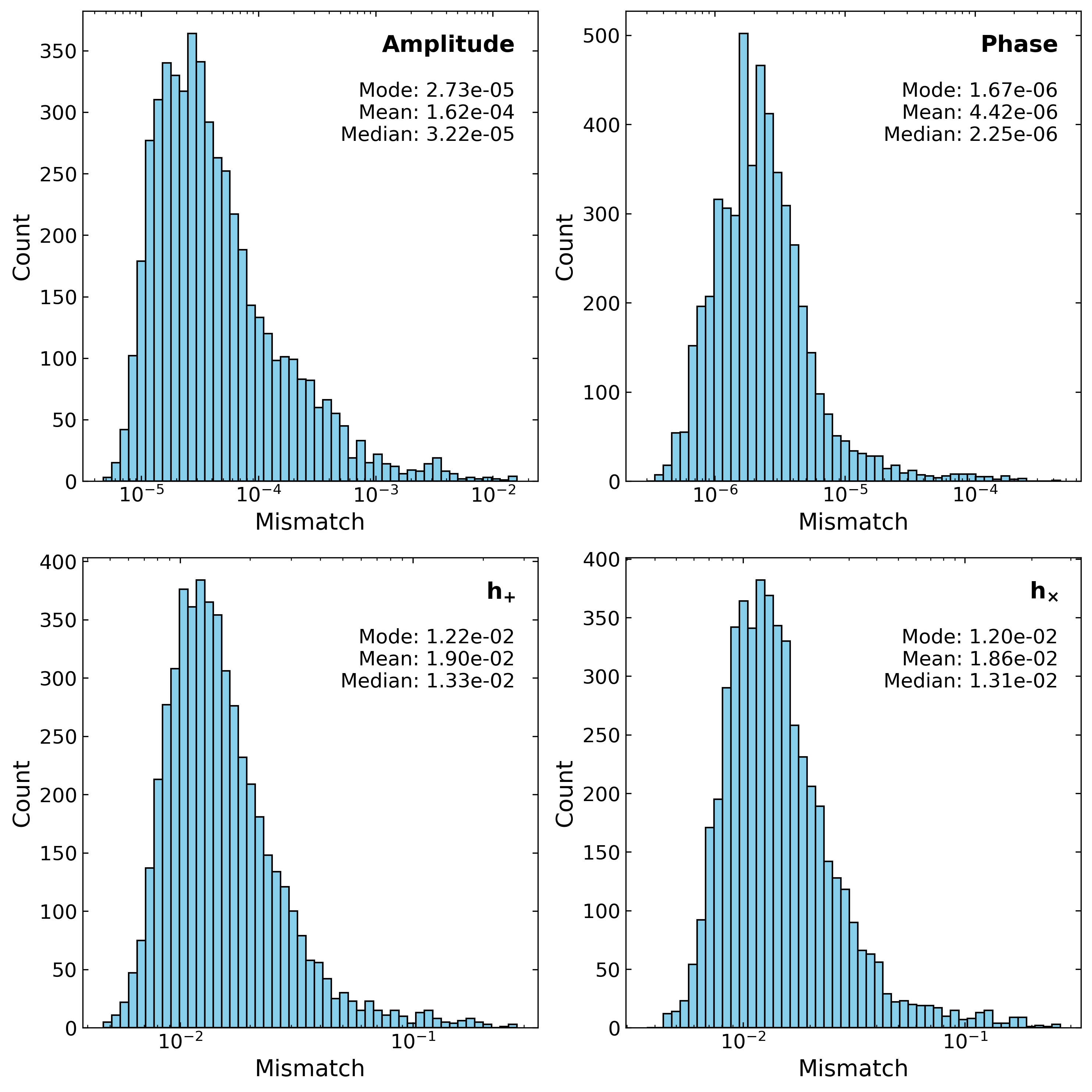}
    \caption{(Top) Distribution of the cosine distance of the \acl{ml} model generated and calibrated amplitude and phase series outputs from the target. (Bottom) Distribution of the mismatch values for the polarization waveforms reconstructed from the calibrated \acl{ml} model outputs. All distributions are shown for data in the test dataset.}
    \label{fig:p2-mm-hist}
\end{figure}


\begin{figure*}
    \centering
    \includegraphics[width=\linewidth]{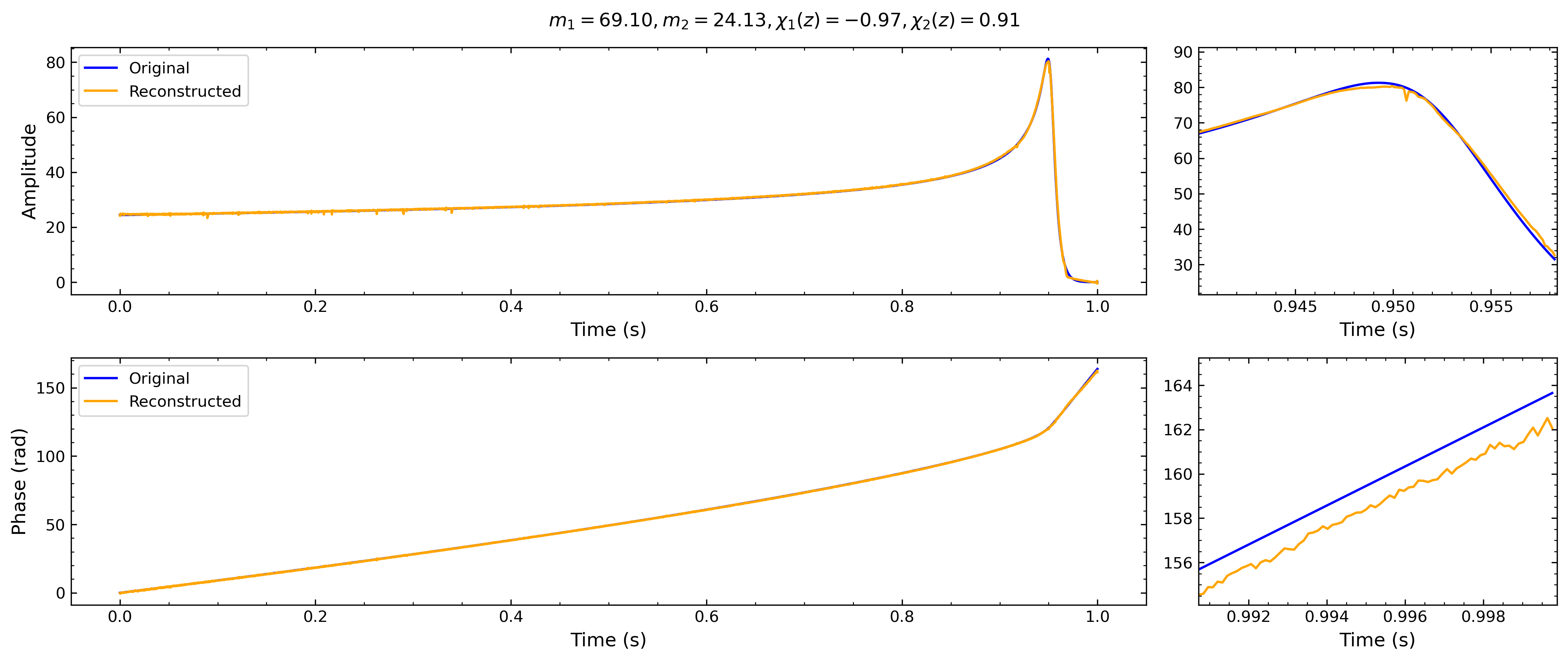}
    \caption{An example overplot of the calibration result for the two-stage waveform generator model. The \acl{ml} model outputs are normalized amplitude and phase.}
    \label{fig:p2-cal-overplot}
\end{figure*}


Similar to \citet{garg2026a}, we evaluate the performance of our \acl{ml} waveform model using a reserved test dataset the model has not yet seen.
The two-stage waveform generator first produces the normalized amplitude and phase evolution using the deterministic \texttt{FlexCAEPhase} model, and then applies a calibration stage to reduce the remaining systematic reconstruction error. The calibration network is trained to learn the residual correction between the original waveform representation and the reconstructed waveform, thereby improving the agreement in both the amplitude and phase channels before converting the result to the gravitational-wave polarizations. \cref{fig:p2-cal-overplot} shows a representative calibration result for normalized \acl{ml} model outputs. The reconstructed amplitude closely follows the original waveform over the full duration, including the inspiral and the rapid merger-ringdown transition. The phase reconstruction is also accurate over most of the signal, with the largest visible deviations appearing only close to the endpoint, where small errors accumulate most strongly because of rapid phase evolution.

The mismatch distributions after calibration are shown in \cref{fig:p2-mm-hist}. 
These distributions span a few orders of magnitude and vary widely across the parameter space. Thus, the mode and the median value of the mismatch distributions are the appropriate measures of the central tendency. We also quote the mean value for completeness' sake.
The calibrated amplitude and phase reconstructions achieve very low cosine distance errors, with median values of $3.22\times10^{-5}$ and $2.25\times10^{-6}$, respectively. The corresponding mean cosine distance between the output and the target are $1.62\times10^{-4}$ for amplitude and $4.42\times10^{-6}$ for phase, indicating that the two-stage model substantially improves the internal amplitude-phase representation. After conversion to the physical polarizations, the median polarization mismatches (calculated using the noise-weighted inner product as in \cite[Equation 7]{garg2026a}) become $1.33\times10^{-2}$ for $h_{+}$ and $1.86\times10^{-2}$ for $h_{\times}$, with mean values of $1.90\times10^{-2}$ and $1.86\times10^{-2}$, respectively. These results show that the calibration stage is effective at reducing waveform-representation errors, although the polarization mismatch remains more sensitive to small residual phase and amplitude errors, particularly near high-amplitude merger region.


\section{Waveform conditioning}
\label{sec:wf-conditioning}

Now, it turns out, directly using the \acl{ml} model output waveforms for parameter estimation tasks and other applications is not a good idea. Crucially, the following should be taken care of:

\begin{itemize}[nosep]
    \item All waveforms output by the \acl{ml} model should ideally have the merger occuring at the same time within the data segment, i.e. to say that the merger time of the \acl{gw} signal should be aligned appropriately.
    \item Waveform amplitude should start and end at zero within the frequency band of interest.
    \item A data segment used for injection in parameter estimation runs should smoothly enter and leave the detector frequency band.
    \item This data segment containing a signal should also be of sufficient duration, so as to mitigate the edge effects occuring during Fourier transformations.
\end{itemize}

\Acl{gw} parameter estimation with gravitational waveform approximations assumes that in the frequency-domain, all frequency-bins are well approximated as independent Gaussian random variables.
The above conditioning requirements for the \acl{ml} output waveforms and injection data segments, ensure that frequency-domain edge effects at the start and end of the signal do not invalidate this assumption, by causing correlations in the frequency-bins. Sufficient data segment duration is required so that the frequency-bins are not corrupted due to these edge effects. 

For our \acl{ml} generated waveforms, we perform the waveform conditioning in the following manner.
Before transforming the reconstructed time-domain polarizations to the frequency domain, each \qty{1}{\second} waveform segment is embedded into a longer fixed-duration data segment. This conditioning step is necessary because the discrete Fourier transform implicitly assumes that the finite time series is periodically continued outside the analysed interval. If the time-domain waveform starts or terminates abruptly at the boundaries of the segment, this periodic continuation introduces artificial discontinuities, which in turn produce spectral leakage and edge artefacts in the Fourier-domain waveform. To reduce these effects, the generated signal is smoothly tapered at both ends and then placed at the centre of an \qty{8}{\second} long zero-padded array. The resulting time series therefore has negligible strain at the boundaries of the full data segment, while preserving the physical \acl{imr} morphology in the central signal region.

Let $h(t)$ denote either of the two polarization waveforms, sampled over a \qty{1}{\second} interval. The conditioned waveform is constructed by multiplying $h(t)$ by a smooth window $w(t)$, where the beginning of the signal is tapered with a $\sin^2$ turn-on and the end of the signal is tapered with a $\cos^2$ turn-off. The start taper extends over one gravitational-wave cycle of duration $T_{\rm cyc}$, and the end taper extends over a duration $T_{\rm end}$, which we choose to be $10\%$ of the post-merger portion of the signal. Schematically, the tapering window can be written as
\begin{equation}
    w(t)=
    \begin{cases}
    \sin^2\left[\dfrac{\pi}{2}\dfrac{t-t_{\rm start}}{T_{\rm cyc}}\right],
    & t_{\rm start}\leq t < t_{\rm start}+T_{\rm cyc}, \\[1.2em]
    1,
    & t_{\rm start}+T_{\rm cyc}\leq t < t_{\rm end}-T_{\rm end}, \\[1.2em]
    \cos^2\left[\dfrac{\pi}{2}\dfrac{t-(t_{\rm end}-T_{\rm end})}{T_{\rm end}}\right],
    & t_{\rm end}-T_{\rm end}\leq t \leq t_{\rm end}.
    \end{cases}
\end{equation}
The tapered waveform is then
\begin{equation}
h_{\rm tap}(t)=w(t)h(t).
\end{equation}
The $\sin^2$ factor ensures that the signal rises smoothly from zero at the beginning of the segment, while the $\cos^2$ factor smoothly suppresses the residual post-merger/ringdown tail towards zero at the end of the waveform. This avoids introducing a sharp jump between the signal and the surrounding zero padding.

\begin{figure}
    \centering
    \includegraphics[width=\linewidth]{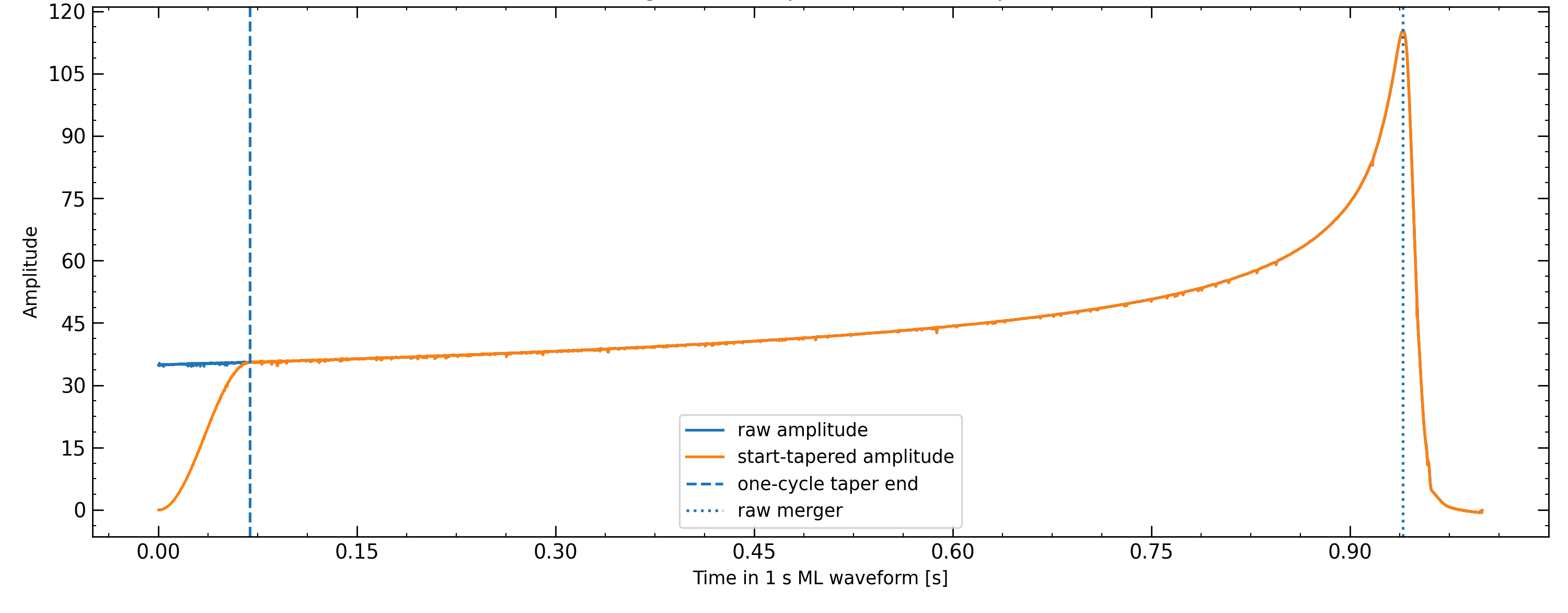}
    \caption{Schematic showing waveform conditioning performed by applying a $\sin^2(t)$ taper at the start of the amplitude series output by the \acl{ml} model.}
    \label{fig:p2m-taper-amp}
\end{figure}

\begin{figure}
    \centering
    \includegraphics[width=\linewidth]{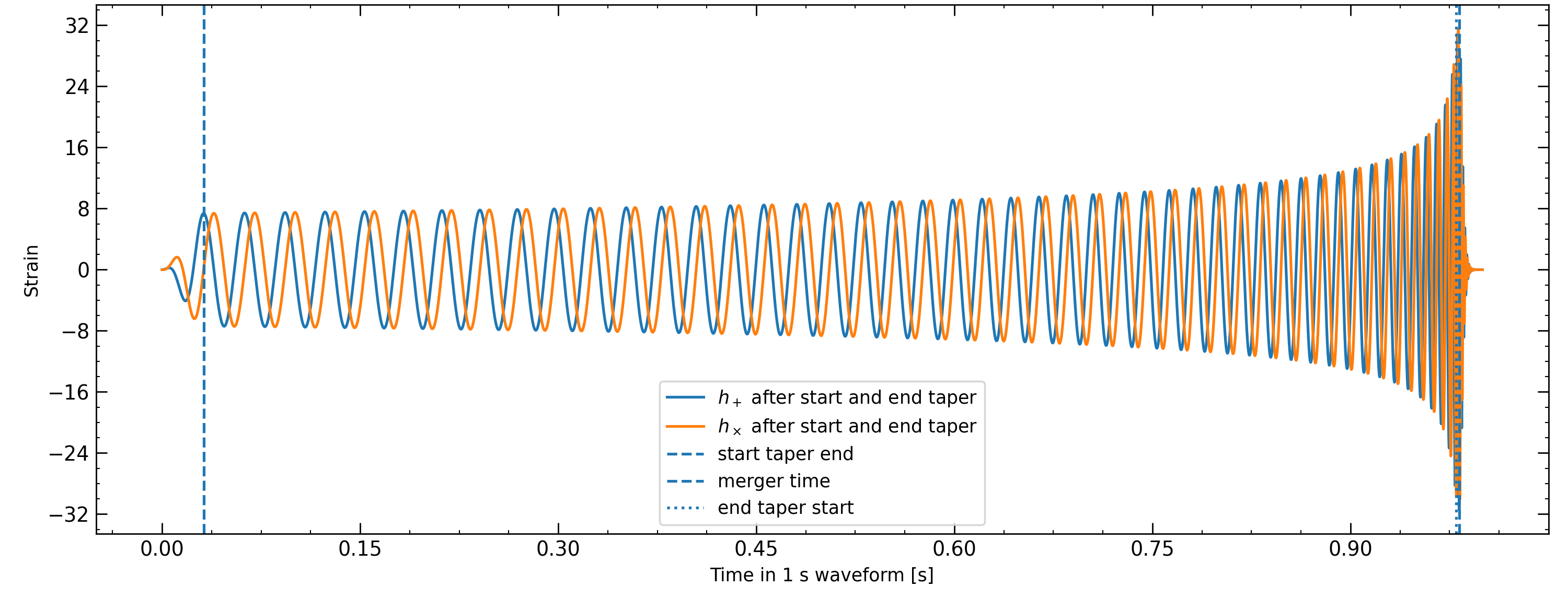}
    \caption{An example overplot of the \acl{gw} polarization strain with the waveform conditioning applied to the outputs of the two-stage waveform generation and calibration model.}
    \label{fig:p2m-cond-overplot}
\end{figure}

After tapering, $h_{\rm tap}(t)$ is inserted into the middle of an \qty{8}{\second} array $H(t)$, with zeros before and after the \qty{1}{\second} signal region. If $T_{\rm seg}=8\,{\rm s}$ and $T_{\rm sig}=1\,{\rm s}$, the signal is placed such that its midpoint coincides with the midpoint of the full segment. Equivalently, the nonzero waveform occupies the interval
\begin{equation}
    \left[
    \frac{T_{\rm seg}-T_{\rm sig}}{2},
    \frac{T_{\rm seg}+T_{\rm sig}}{2}
    \right],
\end{equation}
inside the full data segment. The final conditioned strain time series is therefore
\begin{equation}
    H(t)=
    \begin{cases}
    h_{\rm tap}\left(t-\dfrac{T_{\rm seg}-T_{\rm sig}}{2}\right),
    & \dfrac{T_{\rm seg}-T_{\rm sig}}{2}\leq t
    \leq \dfrac{T_{\rm seg}+T_{\rm sig}}{2}, 
    \\[1.2em]
    0,
    & \text{otherwise}.
    \end{cases}
\end{equation}
This \qty{8}{\second} conditioned time-domain data segment is then Fourier transformed to the frequency-domain, before any use for parameter estimation analysis. The longer segment also sets the discrete frequency spacing to $\Delta f=1/T_{\rm seg}=0.125\,{\rm Hz}$, while the tapering primarily controls the edge behaviour of the finite-duration signal before the Fourier-domain analysis.


\section{Results with conditioned waveforms}

\begin{figure}
    \centering
    \includegraphics[width=\linewidth]{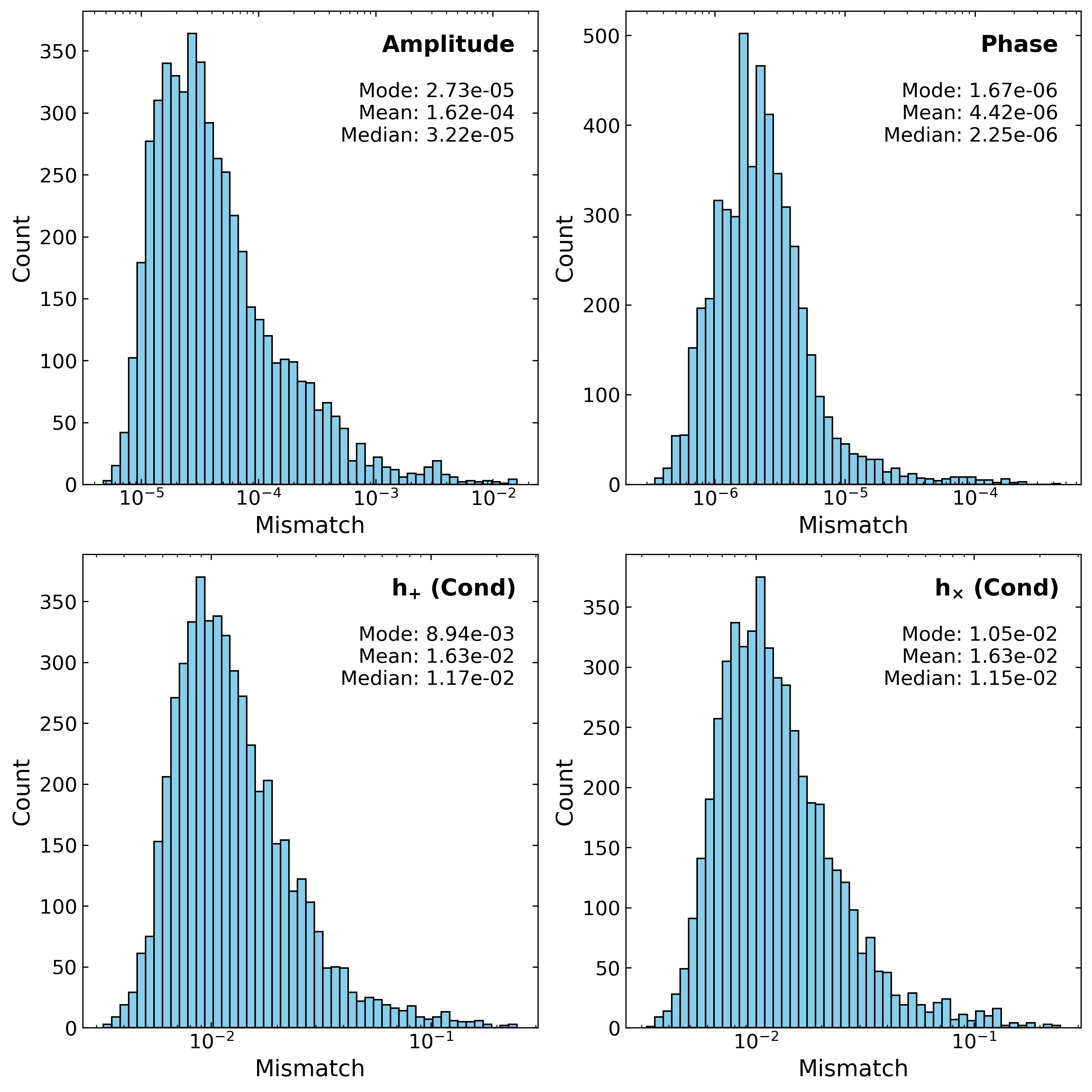}
    \caption{(Top) Distribution of the cosine distance for the conditioned amplitude and phase series outputs from the two-stage waveform generation and calibration model. (Bottom) Distribution of the mismatch values for the polarization waveforms reconstructed from conditioned outputs. All distributions are shown for data in the test dataset.}
    \label{fig:p2m-cond-mmhist}
\end{figure}

\begin{figure}
    \centering
    \includegraphics[width=\linewidth]{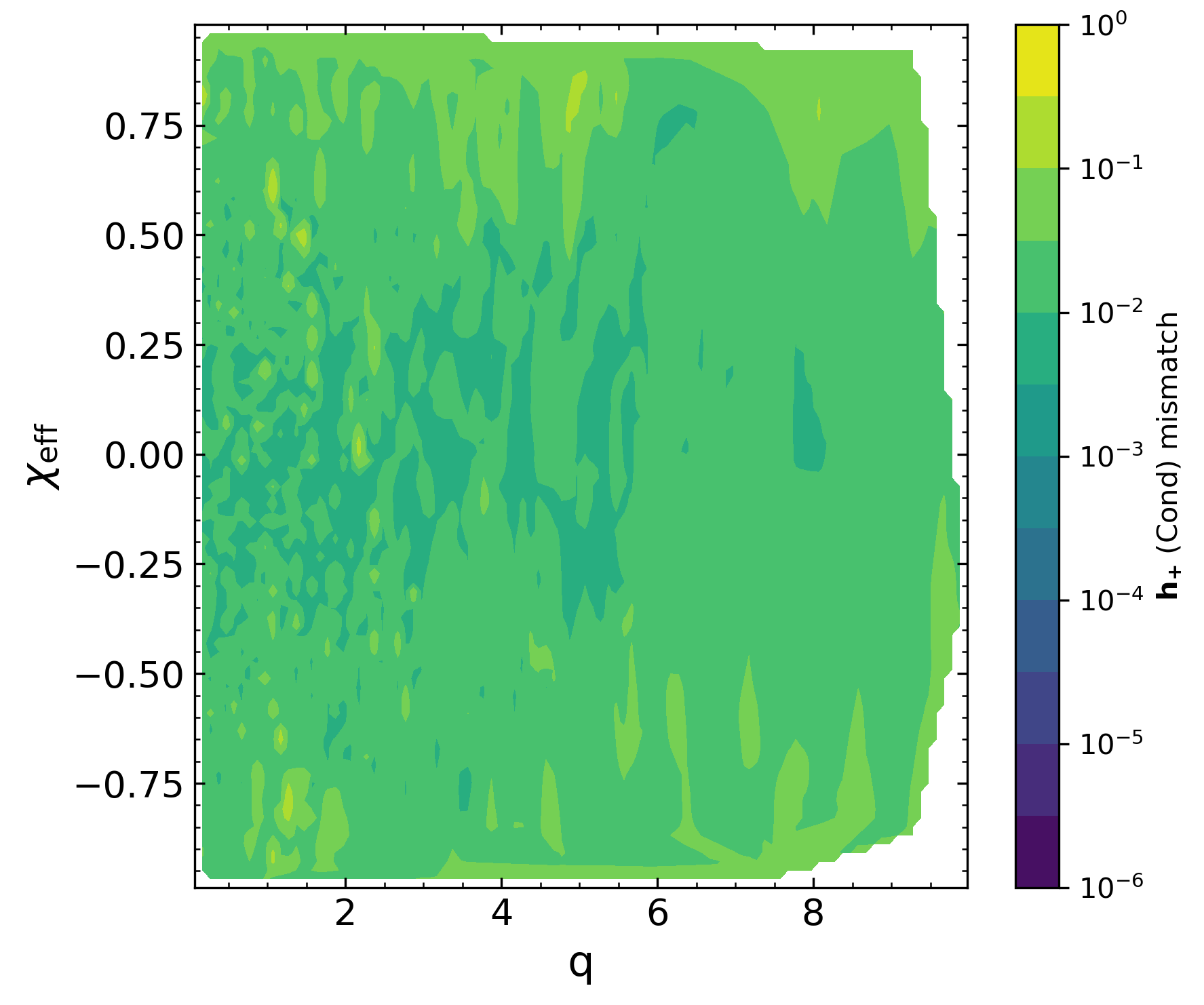}
    \caption{Contour plot of the mismatch values for $h_{+}$ reconstruction from conditioned wavefrom outputs of the two-stage waveform generation and calibration model. Full parameter space is used, i.e. $\chieff\in[-0.99,0.99]$.}
    \label{fig:p2m-cond-contour}
\end{figure}

\begin{figure}
	\centering
	\includegraphics[width=0.5\linewidth]{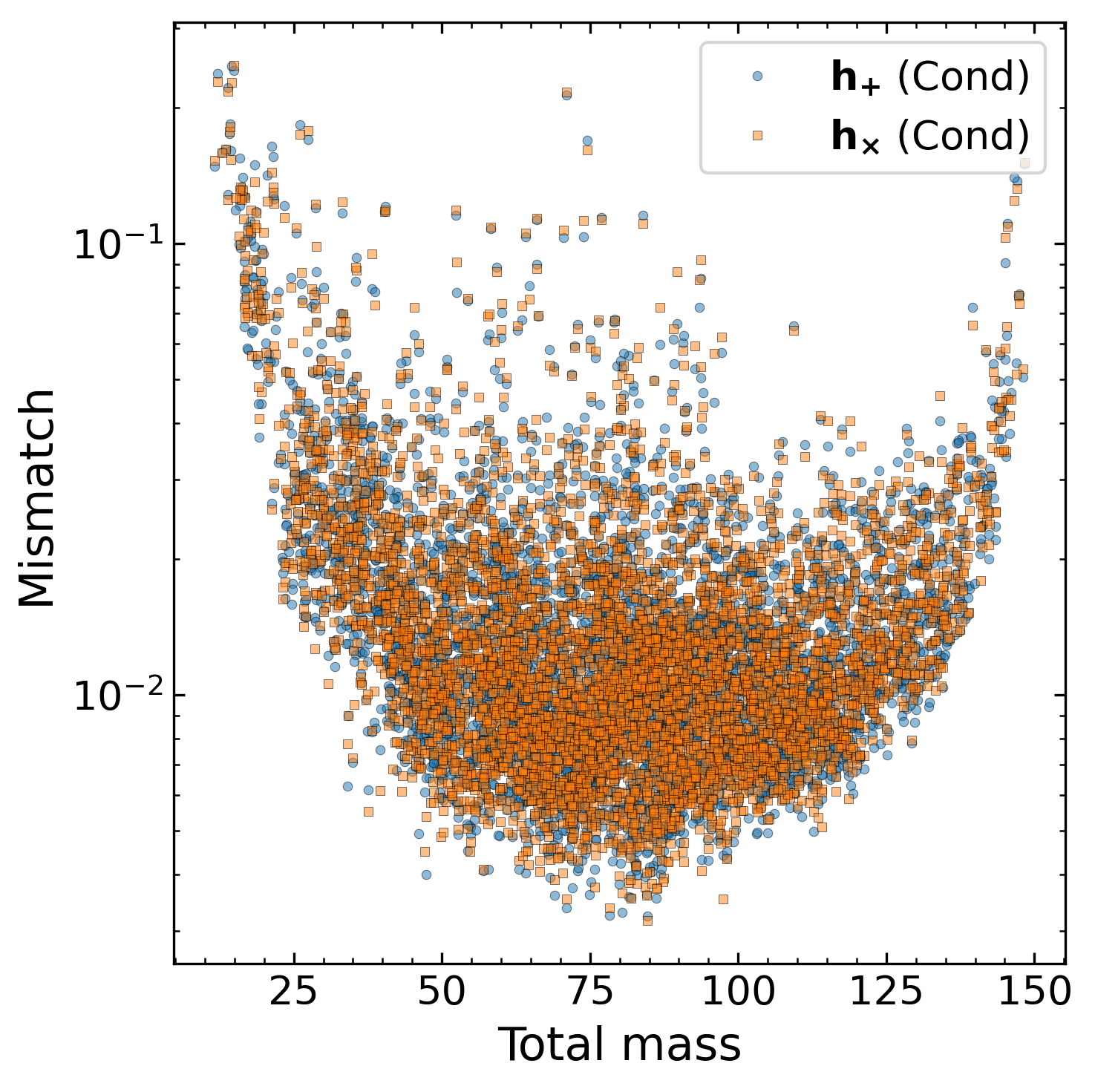}%
	\includegraphics[width=0.5\linewidth]{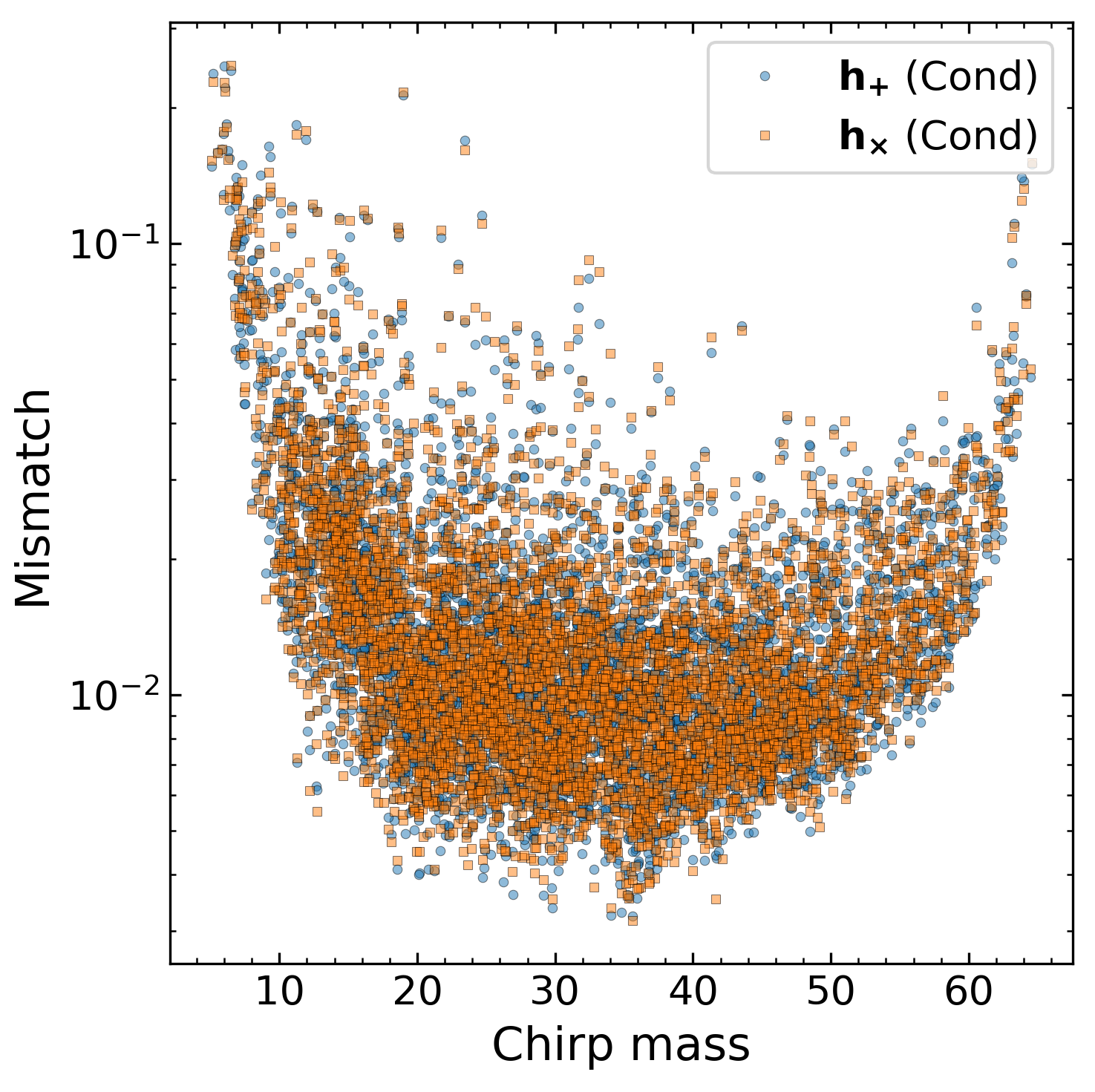}
	\includegraphics[width=0.5\linewidth]{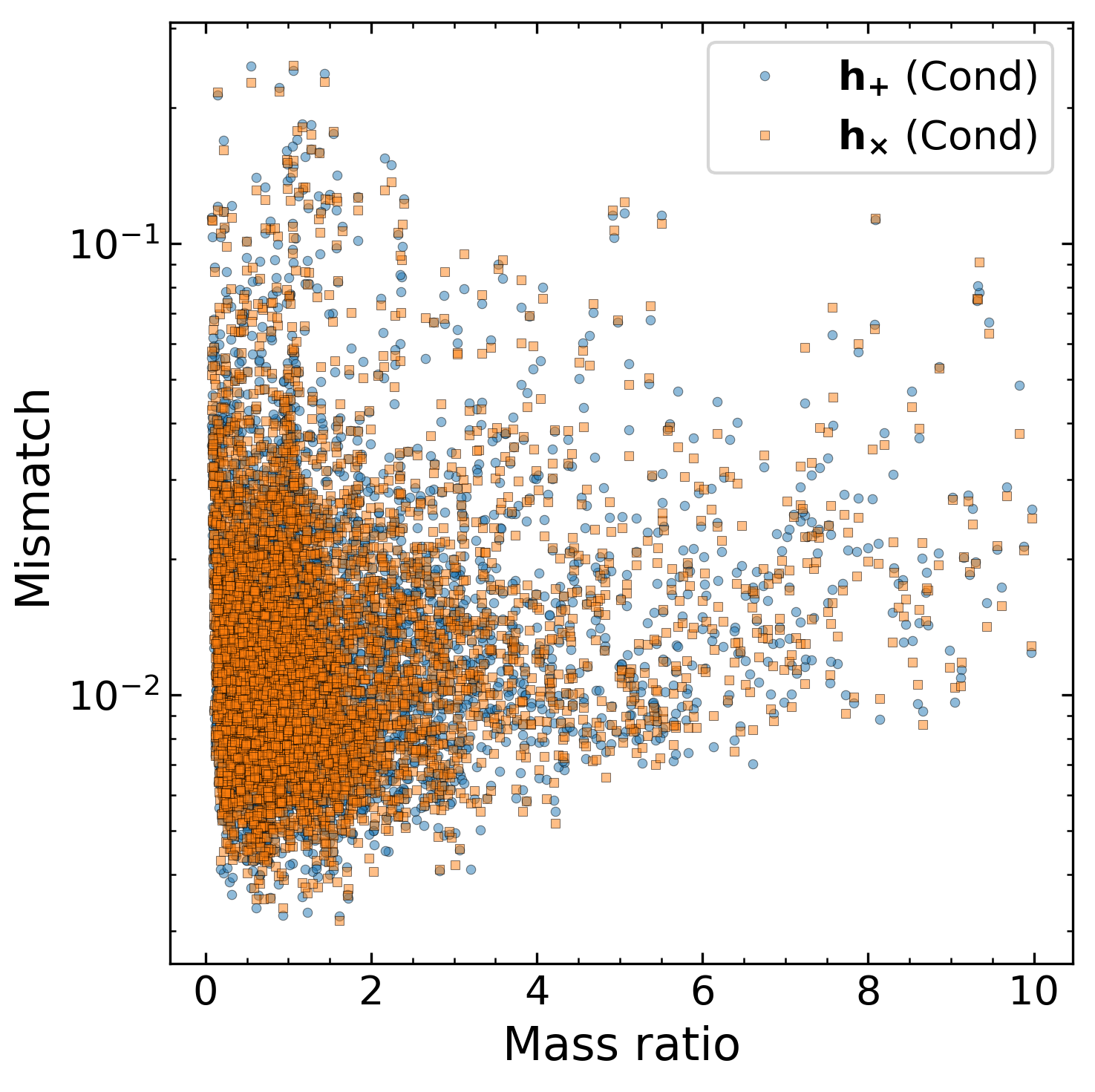}%
	\includegraphics[width=0.5\linewidth]{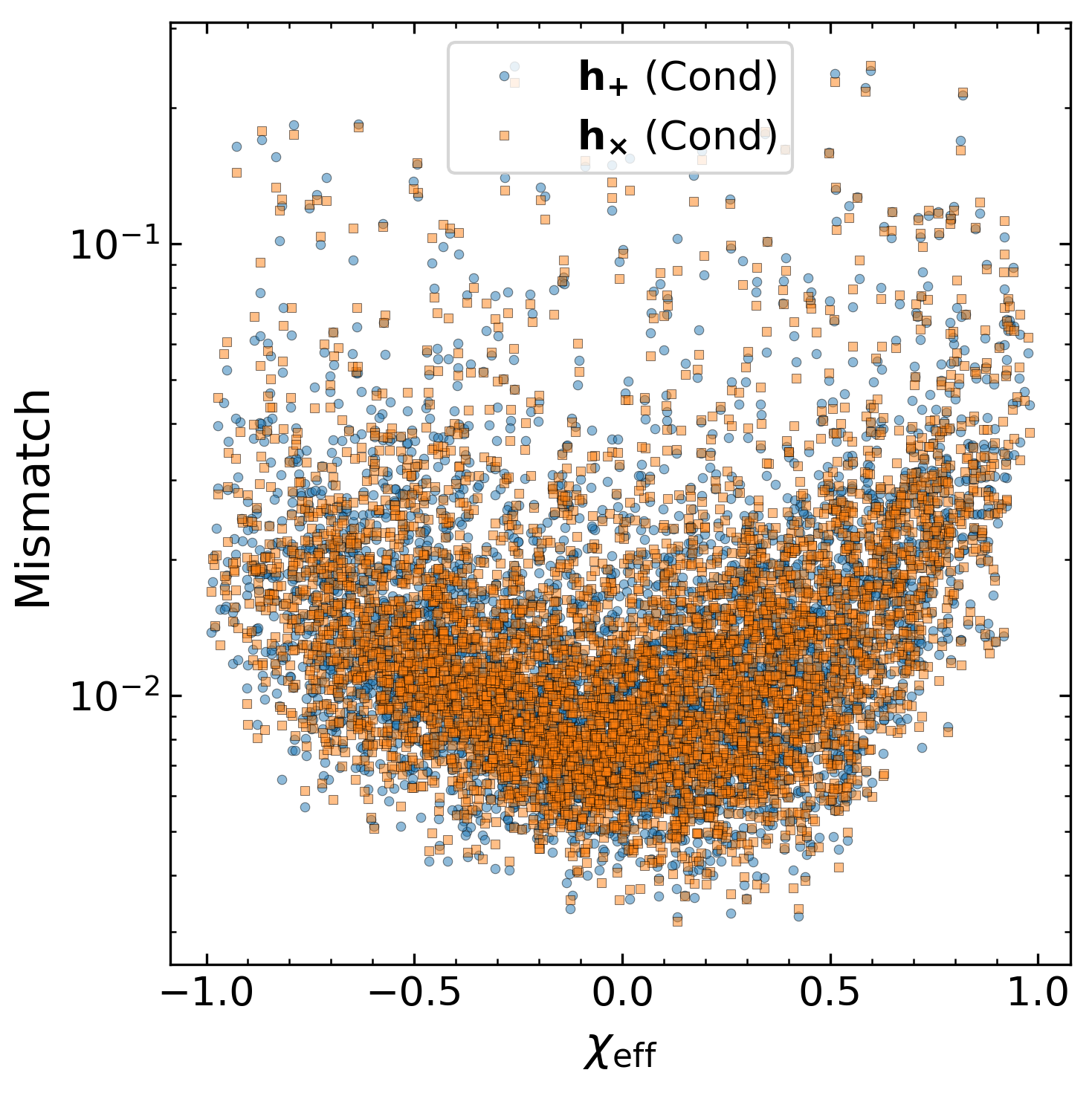}
	\caption{Mismatch values as a function of total mass (top left), chirp mass (top right), mass ratio (bottom left), and effective spin (bottom right) for conditioned waveform outputs from the two-stage waveform generation and calibration model, for all data in the test dataset. Full parameter space is used, i.e. $\chieff\in[-0.99,0.99]$.}
	\label{fig:p2m-cond-mmplot}
\end{figure}

We next evaluate the performance of the conditioned-output waveform-generation model after reconstructing the time-domain waveform representation, applying the start--end tapering procedure and embedding the \qty{1}{\second} waveform into an \qty{8}{\second} zero-padded data segment. The resulting mismatch distributions are shown in \cref{fig:p2m-cond-mmhist}. The amplitude and phase representations are reconstructed with substantially smaller errors than the final polarization waveforms. For the amplitude reconstruction, the histogram-bin mode, mean, and median mismatches are $2.73\times10^{-5}$, $1.62\times10^{-4}$, and $3.22\times10^{-5}$, respectively. For the phase reconstruction, the corresponding values are $1.67\times10^{-6}$, $4.42\times10^{-6}$, and $2.25\times10^{-6}$. These small representation-level mismatches indicate that the conditioned-output model continues to learn the smooth amplitude and phase structure of the waveform accurately. As expected, the accuracy diminishes after reconstructing the oscillatory polarizations $h_+$ and $h_\times$, since small residual errors in amplitude and phase can accumulate to coherent mismatch errors when mapped back to strain. The conditioned polarization waveforms have comparable mismatch distributions for the two polarizations: for $h_+$, the histogram-bin mode, mean, and median are $8.94\times10^{-3}$, $1.63\times10^{-2}$, and $1.17\times10^{-2}$, while for $h_\times$ they are $1.05\times10^{-2}$, $1.63\times10^{-2}$, and $1.15\times10^{-2}$. The close agreement between the $h_+$ and $h_\times$ statistics suggests that the conditioning procedure does not preferentially degrade one polarization relative to the other.

The dependence of the polarization mismatch for conditioned waveforms on the source parameters is shown in \cref{fig:p2m-cond-mmplot}. The scatter plots against chirp mass and total mass show a characteristic U-shaped trend, with the smallest mismatches concentrated in the central part of the mass range and larger mismatches appearing near the lower- and higher-mass boundaries. This behaviour is consistent with the fact that the morphology and duration of the time-domain signal change substantially across the mass range: low-mass systems contain longer inspiral structure within the fixed signal window, while high-mass systems are more merger--ringdown dominated. The mismatch also increases for more asymmetric systems, as seen in the mass-ratio dependence. Most samples remain clustered around a waveform mismatch of $\mismatch\sim10^{-2}$, but the high-mass-ratio region contains a larger fraction of high-mismatch outliers. A similar boundary effect is visible in the effective-spin dependence, where the lowest mismatches are generally obtained near moderate $\chi_{\rm eff}$, while larger errors appear more frequently towards large positive or negative aligned effective spin. 
The two-dimensional mismatch map in the $(q,\chi_{\rm eff})$ plane (\cref{fig:p2m-cond-contour}) shows the same broad trend: the model is most stable over the densely populated central parameter region, while the mismatch becomes more variable near the edges of the training domain.

Overall, the conditioned-output waveform model achieves accurate reconstruction at the amplitude--phase representation level, with typical mismatches between $10^{-6}$ and $10^{-4}$, while the final polarization mismatches are typically of $\order{10^{-2}}$. The conditioning and tapering procedure successfully produces finite-duration waveforms suitable for Fourier-domain analysis, but it does not by itself remove the dominant residual mismatch in the reconstructed polarizations. The persistence of $\order{10^{-2}}$ polarization mismatch, despite the much smaller amplitude and phase representation errors, suggests that the main remaining limitation is the sensitivity of the reconstructed strain to small coherent amplitude--phase residuals, particularly near the boundaries of the physical parameter space. 


\section{Parameter inference with machine learning surrogate: discussion and results}
\label{sec:model2pe}

In this section, we now use the conditioned waveforms outputs from the two-stage waveform generation and calibration model (described in \cref{sec:wf-conditioning}) for \acl{pe} tests.
\Acl{pe} with our \acl{ml} waveforms can be performed by adding a custom waveform Python class to \texttt{Bilby} \cite{ashton2019}.
Due to constaint that the duration of the \acl{ml} generated waveforms is fixed at \qty{1}{\second}, for \acl{pe} analysis we use a restricted mass prior range of $m_1,m_2\in \mathrm{Uniform}[30,75]$, such that the target \ac{eob} waveforms used for likelihood calculation are atleast \qty{1}{\second} duration for a $f_{\rm low}$ of \qty{14}{\hertz}. The two spin components are uniform in the range $[-0.80,0.80]$. All \acl{ml} model generated waveforms are also scaled to be at different luminosity distances as required during the \acl{pe} runs.

Let $C[\cdot]$ denote the conditioning operation, consisting of fixed merger-time alignment, tapering, and embedding in the longer \qty{8}{\second} data segment. The conditioned \acl{ml} waveform is then
\begin{equation}
    h_{\rm ML}^{C}(t;\params)
    =
    C\!\left[h_{\rm ML}(t;\params)\right],
\end{equation}
and the corresponding conditioned \ac{eob} reference waveform is
\begin{equation}
    h_{\rm EOB}^{C}(t;\params)
    =
    C\!\left[h_{\rm EOB}(t;\params)\right].
\end{equation}
In particular, the \ac{eob} injection waveforms are processed using the same merger-time alignment, tapering, and embedding procedure as the \acl{ml} waveforms. This ensures that all comparisons are performed within the waveform domain actually learned by the surrogate model. For the \acl{pe} analysis in this work, we will inject waveforms to Gaussian noise data generated with the \texttt{aLIGOZeroDetunedHighPower} \acl{psd} from \texttt{LalSuite} \cite{LALSuite2018}. For parameter sampling we use the \texttt{nessai} sampler \cite{williams2021a} backend in \texttt{Bilby} in all our analysis. Depending on which waveforms are used for injections and which are used for parameter recovery, we have two different \acl{pe} cases: ML2ML and EOB2ML, while the EOB2EOB is the \acl{pe} without using any \acl{ml} surrogates.

\subsection{Posterior-predictive probability-probability validation}

To assess the statistical calibration of the \acl{pe} pipeline, we use \ac{pp} plots constructed from a set of \acl{pe} analysis results for different simulated injections. For each injection $i$, the true source parameters $\boldsymbol{\params}^{\rm inj}_i$ are known. After performing Bayesian inference, the sampler returns posterior samples from $P(\boldsymbol{\params}\mid d_i)$, where $d_i$ denotes the simulated detector data for the $i$-th injection. For a given parameter $\params$, we compute the posterior credible level at which the injected value is recovered,
\begin{equation}
    c_{i,\params}
    = P(\params < \params^{\rm inj}_i \mid d_i)
    \simeq
    \frac{1}{N_i}\sum_{j=1}^{N_i}
    \mathbf{1}\left(\theta_{ij}<\params^{\rm inj}_i\right),
\end{equation}
where $\theta_{ij}$ denotes the $j$-th posterior sample for parameter $\params$ in injection $i, N_i$ is the number of posterior samples, and $\mathbf{1}(\cdot)$ is the indicator function. 


If the \acl{pe} analysis is unbiased, the probability distribution for the parameters will have the correct coverage, meaning the relative frequency of occurence of the true value of each parameter within some interval will equal the integral of the posterior probability distribution within that interval. If we conduct $N_{\rm inj}$ trials, drawing each test case from the prior distribution density, and select a credible interval, ${c_{i,\params}}_{i=1}^{N_{\rm inj}}$, from the posterior distribution density for some parameter $\params$, then the fraction of test cases whose true values for that parameter are within that credible interval is expected to equal that credible interval's probability. By plotting the fraction of simulations found within a credible interval as a function of the size of the credible interval we can empirically confirm the coverage of the posterior probability distribution function. The definition of the credible interval is irrelevant. Recall that this is called the \ac{pp}-plot, for the following function,
\begin{equation}
    \widehat{F}_{\params}(x)
    =
    \frac{1}{N_{\rm inj}}
    \sum_{i=1}^{N_{\rm inj}}
    \mathbf{1}\left(c_{i,\params}<x\right),
    \qquad 0\le x\le 1.
\end{equation}
In a perfect unbiased analysis, $\widehat{F}_{\params}(x)=x$, so the \ac{pp} curve follows the diagonal line. Deviations from the diagonal indicate imperfect posterior calibration. For example, systematic deviations may indicate biased parameter recovery, while overly steep or shallow curves can indicate posterior intervals that are respectively too broad or too narrow. Because of the finite number of trials, random deviations from the diagonal are expected, but we can predict the size of the deviations that we expect using the binomial distribution.

In this work, all \ac{pp} plots are generated using the \texttt{Bilby} result utilities. The horizontal axis denotes the nominal credible interval $x$, while the vertical axis denotes the fraction of injections whose true parameter value lies below that posterior credible level. For each parameter, \texttt{Bilby} computes the injected credible levels and applies a one-sample Kolmogorov-Smirnov test against the uniform distribution on $[0,1]$. The Kolmogorov-Smirnov statistic is
\begin{equation}
    D_{\params}
    =
    \sup_{0\le x\le 1}
    \left|
    \widehat{F}_{\params}(x)-x
    \right|,
\end{equation}
and the associated p-value quantifies the probability, under the null hypothesis of calibrated posteriors, of obtaining a deviation at least as large as the observed one. Small p-values therefore indicate that the credible levels are unlikely to have been drawn from a uniform distribution, suggesting a calibration failure or a systematic bias in the inference setup.

When multiple parameters are tested, \texttt{Bilby} also reports a combined p-value obtained from the individual parameter-wise p-values. This provides a compact diagnostic of the global \ac{pp}-plot consistency. However, since source parameters such as $m_1, m_2, \chi_1$, and $\chi_2$ are generally correlated, the combined p-value should be interpreted as a useful summary statistic rather than as a fully independent hypothesis test. 

In the present analysis, the \ac{pp}-plot is used primarily as a posterior-level self-consistency test of the waveform model and inference pipeline. For the ML2ML case, a diagonal \ac{pp}-plot indicates that injections generated with the \acl{ml} waveform model are statistically recovered by the same model. In contrast, systematic deviations in the EOB2ML case would quantify the impact of waveform-systematic differences between the \ac{eob} injection waveforms and the \ac{ml} recovery waveforms.

\subsection{ML2ML parameter inference with conditioned waveforms}

We first do a \acl{pe} run with likelihood evaluations using \acl{ml} generated waveforms for \acl{ml} waveform injections. These \acl{pe} runs act as a validator for the use of \acl{ml} waveforms with \acl{pe} pipeline, and isolate the parameter sampling and inference procedures from \acl{ml} waveform accuracy errors. This ML2ML case is ideally expected to work well, since whatever accuracy error the \acl{ml} generated waveforms may have, remains identical for both the injections and for the waveforms against which the likelihood is evaluated in this case. Besides, as opposed to the stochastic output of the variational model in \citet{garg2026a}, the outputs of our \acl{ml} waveform model are fully-deterministic, and thus, lead to the same likelihood value if the same set of parameter is sampled twice. 

For the ML2ML \acl{pe} case, the injected signal is generated using the conditioned \ac{ml} waveform model as,
\begin{equation}
    d_{\rm ML}(t) 
    = h_{\rm ML}^{C}(t;\params_{\rm inj})
    + n(t),
\end{equation}
where $n(t)$ is a Gaussian noise realization and $\params_{\rm inj}$ denotes the injected parameter values. The same conditioned waveform model is then used for likelihood calculations. The \ac{ml} likelihood is written schematically as
\begin{multline}
    \log \mathcal{L}_{\rm ML}
    \left(d_{\rm ML}\mid \params\right)
    \\
    = -\frac{1}{2}
    \left(
    d_{\rm ML} - h_{\rm ML}^{C}(\params)
    \,\middle|\,
    d_{\rm ML} - h_{\rm ML}^{C}(\params)
    \right)
    + {\rm const.},
\end{multline}
where $(a|b)$ denotes the usual detector-noise-weighted inner product.
Since the same waveform family is used for injection and recovery, the likelihood is expected to peak near the injected parameters, up to statistical fluctuations due to detector noise. The corresponding posterior is
\begin{equation}
    P_{\rm ML}(\params|d_{\rm ML})
    =
    \frac{
    \pi(\params)\like_{\rm ML}(d_{\rm ML}|\params)
    }{
    Z_{\rm ML}
    },
\end{equation}
where $\pi(\params)$ is the prior and $Z_{\rm ML}$ is the evidence.

Now, recall that for each injection $k$, we compute the percentile value
\begin{equation}
    p_{k}^{(\theta)}
    = \int_{-\infty}^{\theta_{{\rm inj},k}}
    P_{\rm ML}
    \left(
    \theta \mid d_{{\rm ML},k}
    \right)
    \,d\theta
\end{equation}
for each parameter $\theta \in [m_1,m_2,\chi_1,\chi_2]$. If the posterior is calibrated, the set of percentile values $\{p_k^{(\theta)}\}$ should be uniformly distributed on $[0,1]$. Therefore, the \ac{pp}-plot should follow the diagonal within the expected binomial confidence region.

\begin{figure}
	\centering
	\includegraphics[width=\linewidth]{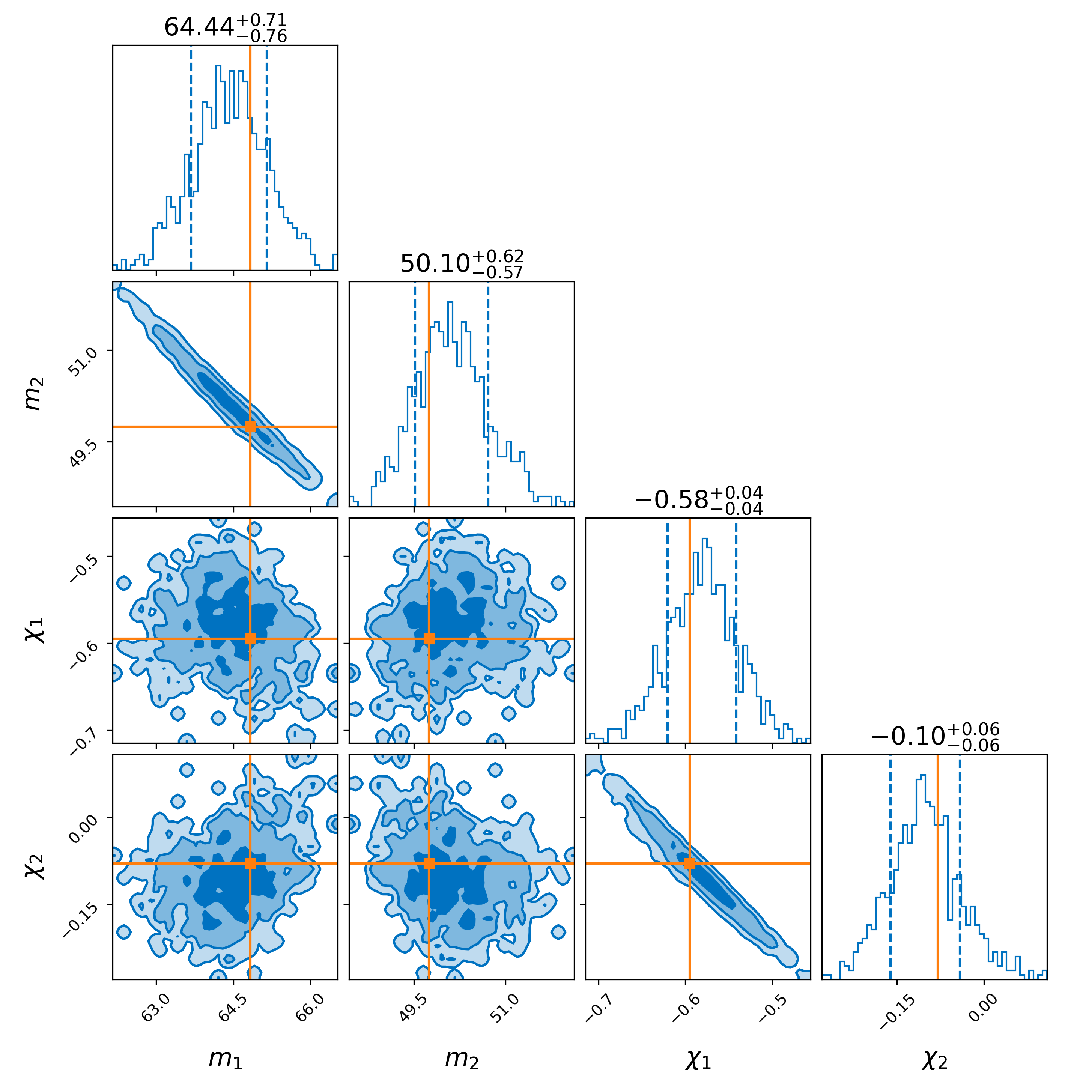}
	\caption{Posterior distributions recovered for a ML2ML \acl{pe} run with conditioned waveform outputs. The injection parameters are $m_1=64.82\Msun$, $m_2=49.75\Msun$, $\chi_1(z)=-0.595$, $\chi_2(z)=-0.08$, $d_L=\qty{400}{\Mpc}$. \Ac{snr} of the injected signal is about $107.12$.
    Other parameters are fixed at: $\theta_{\rm jn}=0.4, \psi=2.659, \phi_c=1.3, t_c=1126259642.413, \mathrm{ra}=1.375$, and $\mathrm{dec}=-1.2108$. The marginalized posterior histograms show the probability density for different posterior samples, alongwith the $1\sigma$ credible interval. The 2D joint posteriors show the 2D $1\sigma, 2\sigma$ and $3\sigma$ credible regions. Each column header denotes the median value of the parameter and the $1\sigma$ credible range around it.
    }
	\label{fig:ml2ml-cond-inj7}
\end{figure}

\begin{figure}
    \centering
    \includegraphics[width=\linewidth]{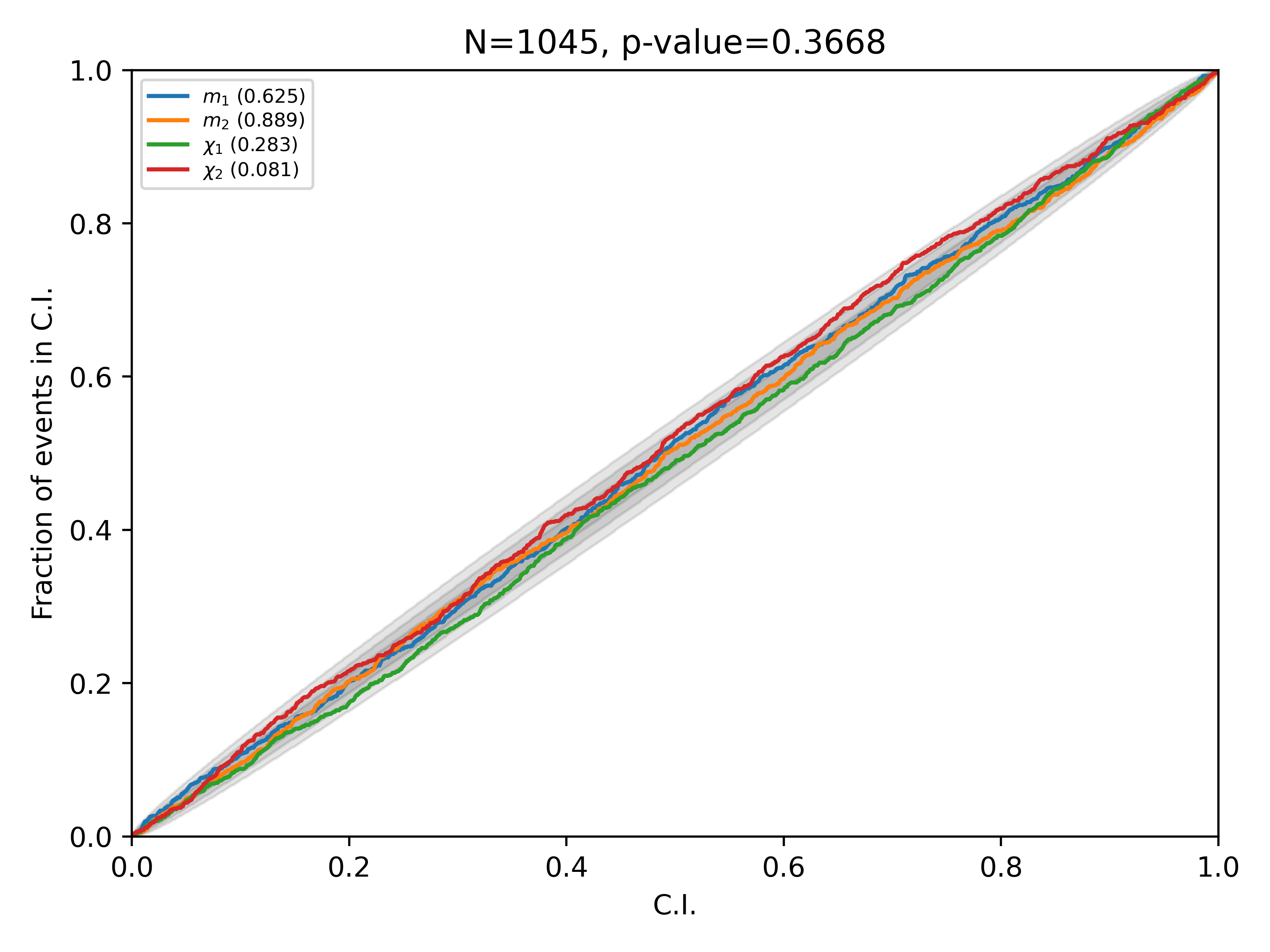}
    \caption{\ac{pp}-plot for the ML2ML recovery experiment with 1045 injections using conditioned waveforms. The luminosity distance for the injections is set at $d_L=\qty{400}{\Mpc}$. The diagonal curve points that there is no systematic error in the \acl{pe} set-up now. All other injection parameters values and other \acl{pe} settings remain the same as in \cref{fig:ml2ml-cond-inj7}.}
    \label{fig:ml2ml-cond-pp}
\end{figure}

\cref{fig:ml2ml-cond-inj7} shows the posterior for one randomly selected injection for the ML2ML \acl{pe} case, while \cref{fig:ml2ml-cond-pp} shows the corresponding \ac{pp}-plot. 
All \acl{pe} runs are set to terminate when the threshold of $\dd\log Z=0.2$ is reached, whereas a lower threshold of around $0.05$ maybe recommended to allow the sampler to contrain a narrower posterior region around the injection value.
We observe that the ML2ML \ac{pp}-plot obtained using the conditioned waveform outputs is consistent with the diagonal. This confirms that the \acl{pe} pipeline, prior implementation, mass ordering, time alignment, waveform conditioning, and likelihood evaluation are internally self-consistent when the injection and recovery waveform families are the same.
In particular, the ML2ML result shows that our \acl{pe} current implementation is capable of recovering injected signals when the data-generating and recovery waveform families are identical.
Therefore, any systematic deviation observed in the EOB2ML case occuring now, cannot be attributed solely to the sampler or to the basic \texttt{Bilby} likelihood calculation setup.


\subsection{EOB2ML parameter inference with conditioned waveforms}

We next perform the EOB2ML \acl{pe} tests. In this case, the injected data are generated using the conditioned \ac{eob} waveform,
\begin{equation}
    d_{\rm EOB}(t)
    = h_{\rm EOB}^{C}(t;\boldsymbol{\theta}_{\rm inj})
    + n(t),
\end{equation}
while the recovery likelihood is still evaluated using the conditioned \acl{ml} waveform model,
\begin{multline}
    \log \mathcal{L}_{\rm ML}
    \left(d_{\rm EOB}\mid \boldsymbol{\theta}\right)
    \\
    = -\frac{1}{2}
    \left(
    d_{\rm EOB}
    - h_{\rm ML}^{C}(\boldsymbol{\theta})
    \,\middle|\,
    d_{\rm EOB}
    - h_{\rm ML}^{C}(\boldsymbol{\theta})
    \right)
    + {\rm const.}
\end{multline}
The corresponding EOB2ML posterior is
\begin{equation}
    P_{\rm EOB2ML}
    \left(
    \boldsymbol{\theta} \mid d_{\rm EOB}
    \right)
    =
    \frac{
    \pi(\boldsymbol{\theta})
    \mathcal{L}_{\rm ML}
    \left(d_{\rm EOB}\mid\boldsymbol{\theta}\right)
    }{
    Z_{\rm EOB2ML}
    }.
\end{equation}%
Now, the EOB2ML likelihood may not necessarily be maximized at $\theta_{\rm inj}$, because the waveform family used for recovery differs from the waveform family used to generate the data. The sampler therefore searches for the parameter value $\theta_{\rm ML}^{\star}$ for which the \ac{ml} waveform best approximates the \ac{eob} injection:
\begin{equation}
    \theta_{\rm ML}^{\star}
    = \arg\max_{\params}
    \like_{\rm ML}(d_{\rm EOB}|\params).
\end{equation}
If $h_{\rm ML}(\theta_{\rm inj})$ differs appreciably from $h_{\rm EOB}(\theta_{\rm inj})$, the maximum-likelihood point $\theta_{\rm ML}^{\star}$ can be systematically displaced from $\theta_{\rm inj}$. This displacement will a waveform-systematic bias rather than arising from any discrepancies in the \acl{pe} or sampler setups, provided that the ML2ML \acl{pe} case is properly calibrated (has a diagonal \ac{pp}-plot), as we have already seen.


The importance of the waveform-systematic bias, inherent in the \ac{pe} process, depends on both the waveform error and the \ac{snr} ratio. For a waveform mismatch $\mismatch$, the likelihood penalty induced by waveform error scales approximately as
\begin{equation}
    \Delta \log \like
    \sim
    \snr^2 \, \mismatch,
    \label{eq:rho2_mismatch_scaling}
\end{equation}
where $\snr$ is the network \ac{snr} ratio. Thus, even a mismatch that appears small in a waveform-reconstruction metric can produce a large posterior bias at high \ac{snr}. A rough condition for waveform errors to be subdominant to statistical uncertainty is
\begin{equation}
    \snr^2 \, \mismatch \lesssim 1.
    \label{eq:mismatch_snr_condition}
\end{equation}
For example, if $\mismatch\sim 10^{-2}$, this condition requires $\snr\lesssim 10$. If $\mismatch\sim 10^{-4}$, the same condition allows $\snr\lesssim 100$. This illustrates why waveform accuracy requirements become more stringent as the signal becomes louder.

\begin{figure}
	\centering
	\includegraphics[width=\linewidth]{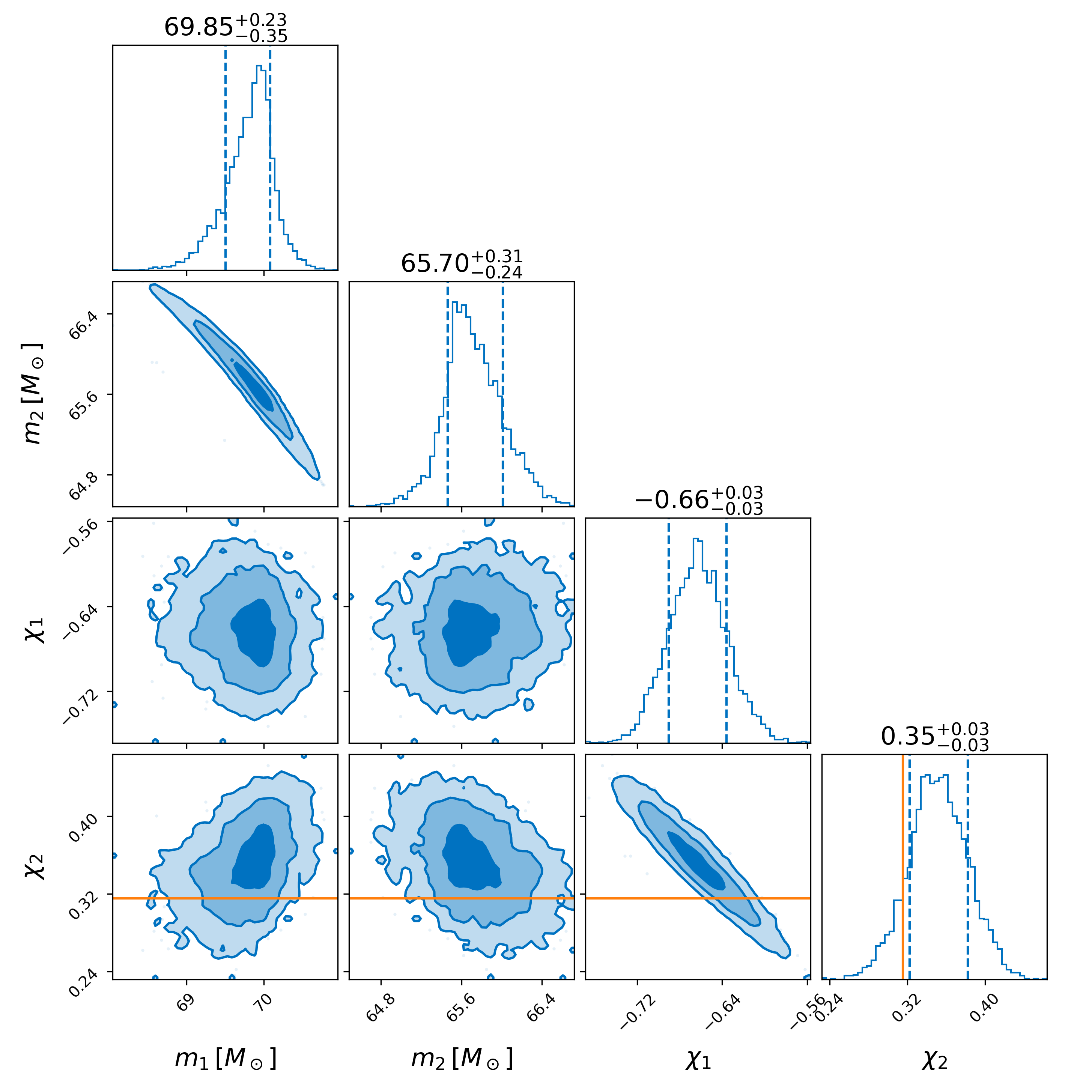}
	\caption{Posterior distributions recovered for a EOB2ML \acl{pe} run with conditioned waveform outputs. The injection parameters are $m_1=74.09\Msun$, $m_2=70.31\Msun$, $\chi_1(z)=-0.32$, $\chi_2(z)=0.31$, $d_L=\qty{400}{\Mpc}$. \Ac{snr} of the injected signal is about $139.45$. Other parameters and \acl{pe} settings remain the same as in \cref{fig:ml2ml-cond-inj7}.}
	\label{fig:eob2ml-cond-inj500}
\end{figure}

\begin{figure}
    \centering
    \includegraphics[width=\linewidth]{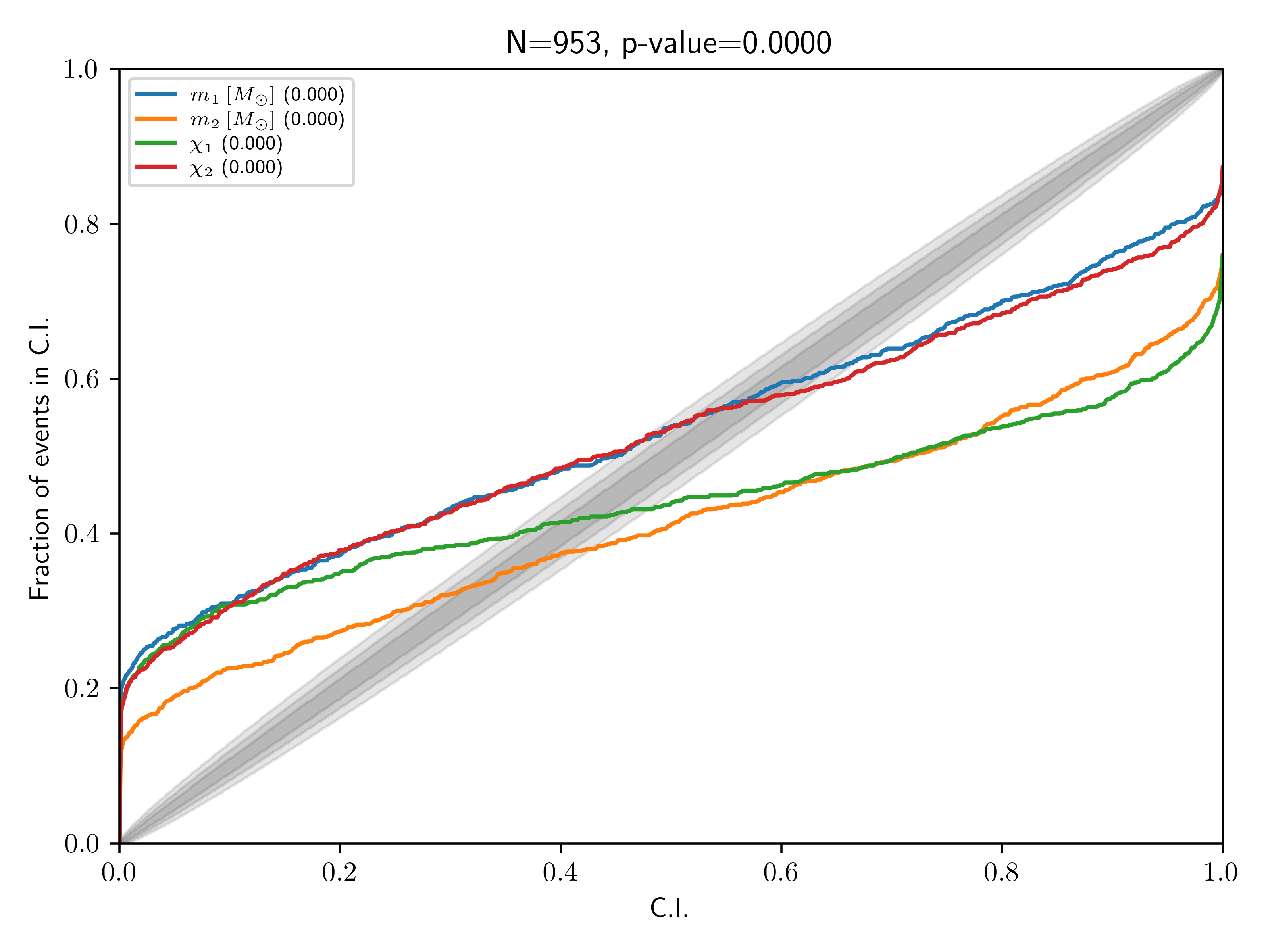}
    \caption{\ac{pp}-plot for the EOB2ML \acl{pe} experiment with 953 injections using conditioned waveforms. Some inherent systematic bias is seen. The luminosity distance for the injections is set at $d_L=\qty{400}{\Mpc}$. All other injection parameters values (apart from the four parameter shown) and other \acl{pe} settings remain the same as in \cref{fig:ml2ml-cond-inj7}.}
    \label{fig:eob2ml-cond-pp}
\end{figure}

\begin{figure}
	\centering
	\includegraphics[width=\linewidth]{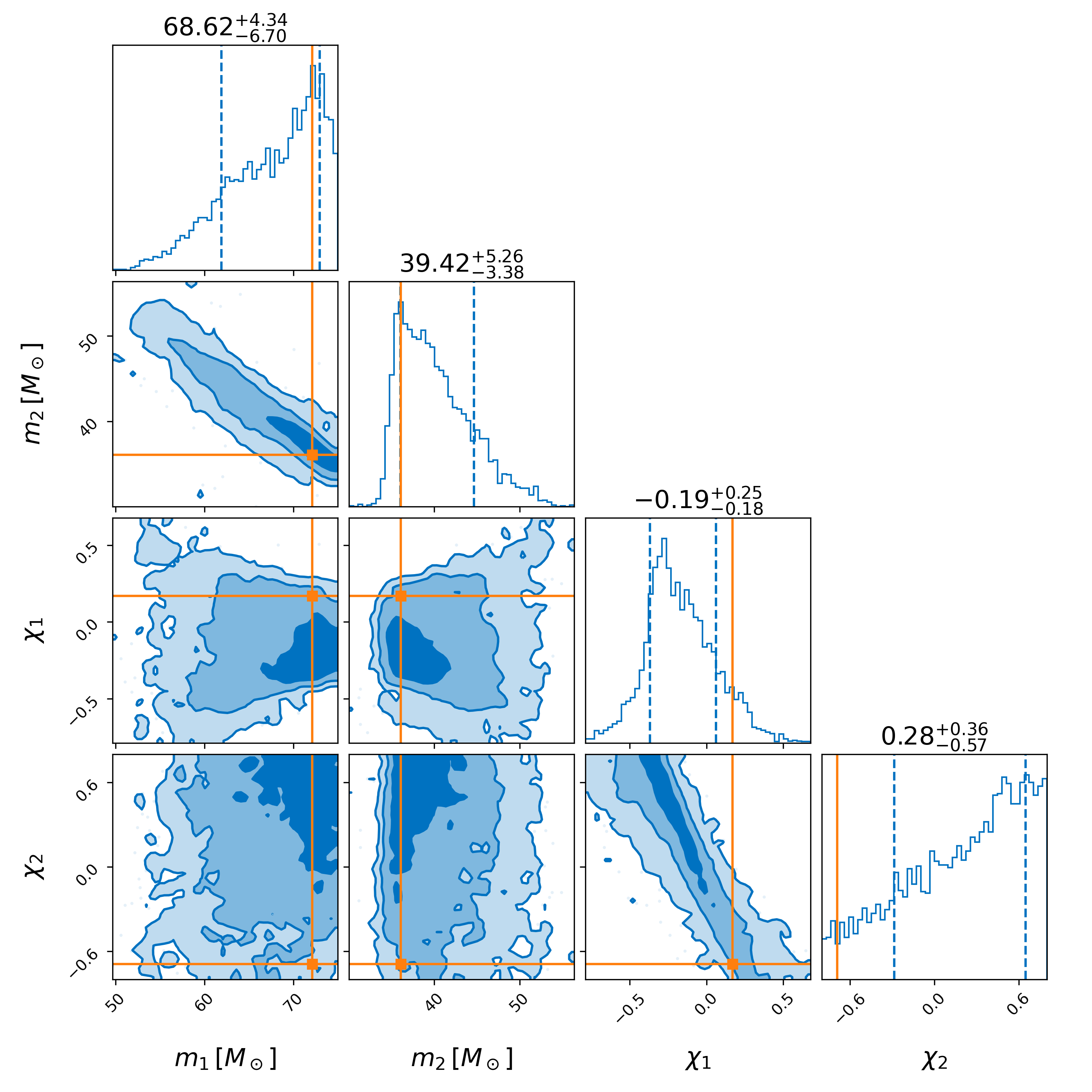}
	\caption{Posterior distributions recovered for a EOB2ML \acl{pe} run with conditioned waveform outputs. The injection parameters are $m_1=72.12\Msun$, $m_2=36.08\Msun$, $\chi_1(z)=0.17$, $\chi_2(z)=-0.69$, $d_L=\qty{4000}{\Mpc}$. \Ac{snr} of the injected signal is about $9.95$. Other parameters and \acl{pe} settings remain the same as in \cref{fig:ml2ml-cond-inj7}.}
	\label{fig:eob2ml-cond-d4k-inj200}
\end{figure}

\begin{figure}
    \centering
    \includegraphics[width=\linewidth]{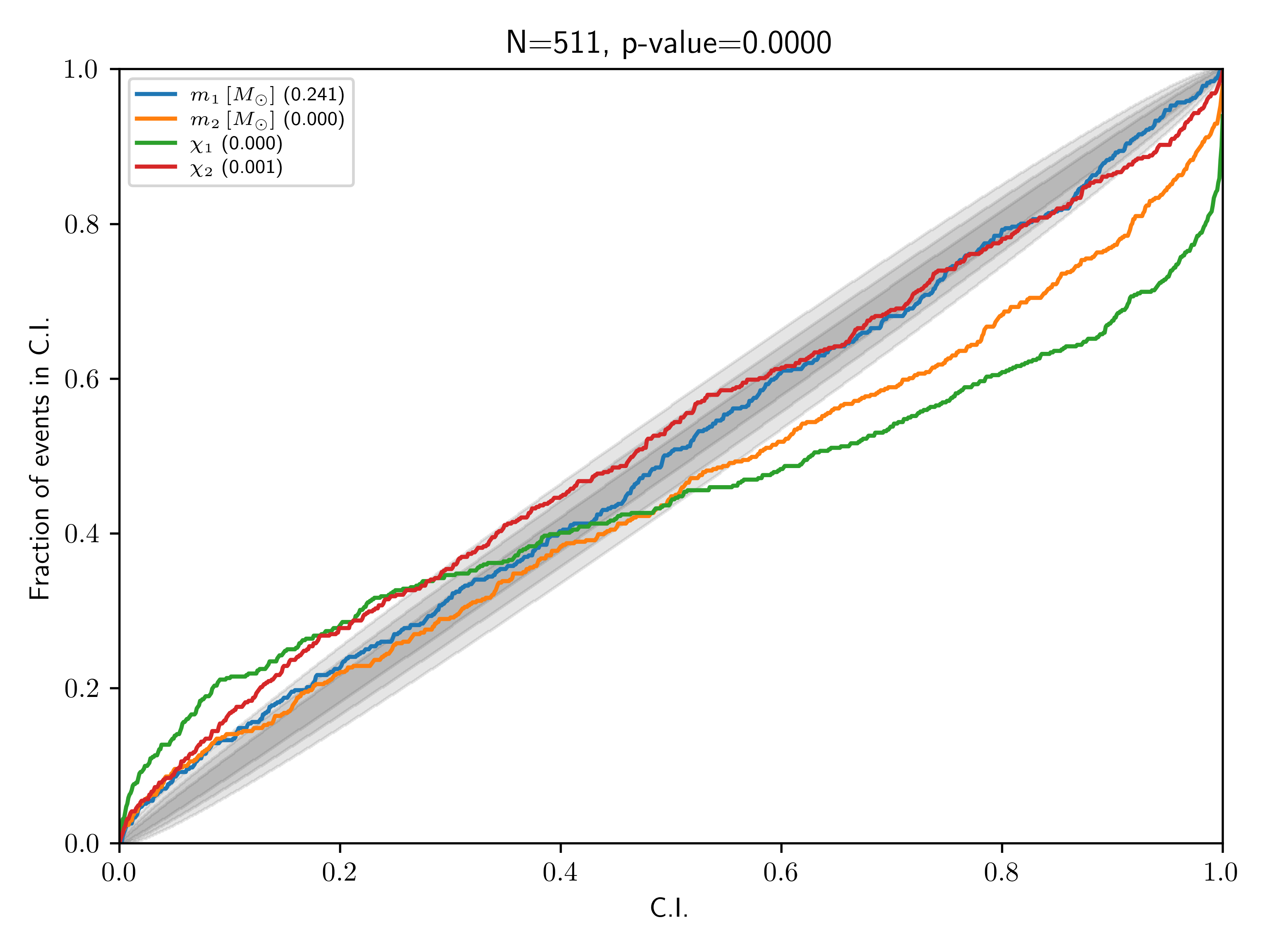}
    \caption{\ac{pp}-plot for the EOB2ML \acl{pe} experiment with 511 injections using conditioned waveforms. The luminosity distance for the injections is set at $d_L=\qty{4000}{\Mpc}$. The \ac{snr} of the signal scales inversely with the luminosity distance to the source. Thus, in comparison to \cref{fig:eob2ml-cond-pp}, we find that there is some improvements in the inherent systematic bias at lower \ac{snr}. All other injection parameters values (apart from the four parameter shown) and other \acl{pe} settings remain the same as in \cref{fig:ml2ml-cond-inj7}.}
    \label{fig:eob2ml-cond-d4k-pp}
\end{figure}

In contrast to the ML2ML case, the EOB2ML posteriors (e.g. \cref{fig:eob2ml-cond-inj500,fig:eob2ml-cond-d4k-inj200}) show systematic deviations from the injected values for a subset of injections. Direct likelihood scans for the EOB2ML posterior demonstrate that the \ac{ml} likelihood evaluated on \ac{eob} injections is often larger at shifted parameter values than at the true injected parameters. The resulting \ac{pp}-plots are not perfectly diagonal and show a characteristic non-uniform structure (\cref{fig:eob2ml-cond-pp,fig:eob2ml-cond-d4k-pp}).
This indicates that, although the conditioned \acl{ml} waveform model gives a self-consistent likelihood when used for both injection and recovery, it does not always reproduce the conditioned \ac{eob} likelihood surface with the required accuracy. The observed deviation can therefore be interpreted as a waveform-family mismatch between $h_{\rm EOB}^{C}(\boldsymbol{\theta})$ and $h_{\rm ML}^{C}(\boldsymbol{\theta})$ under otherwise identical data-analysis conventions.

For a given injection, the EOB2ML posterior may still contain the injected value within a one-dimensional credible interval for some parameters, even when the full four-dimensional posterior is biased. Therefore, we should quantify the inherred bias directly using the inferred one-dimensional posterior summaries.

\subsection{Bias in EOB2ML parameter inference}

\begin{figure}
    \centering
    \includegraphics[width=\linewidth]{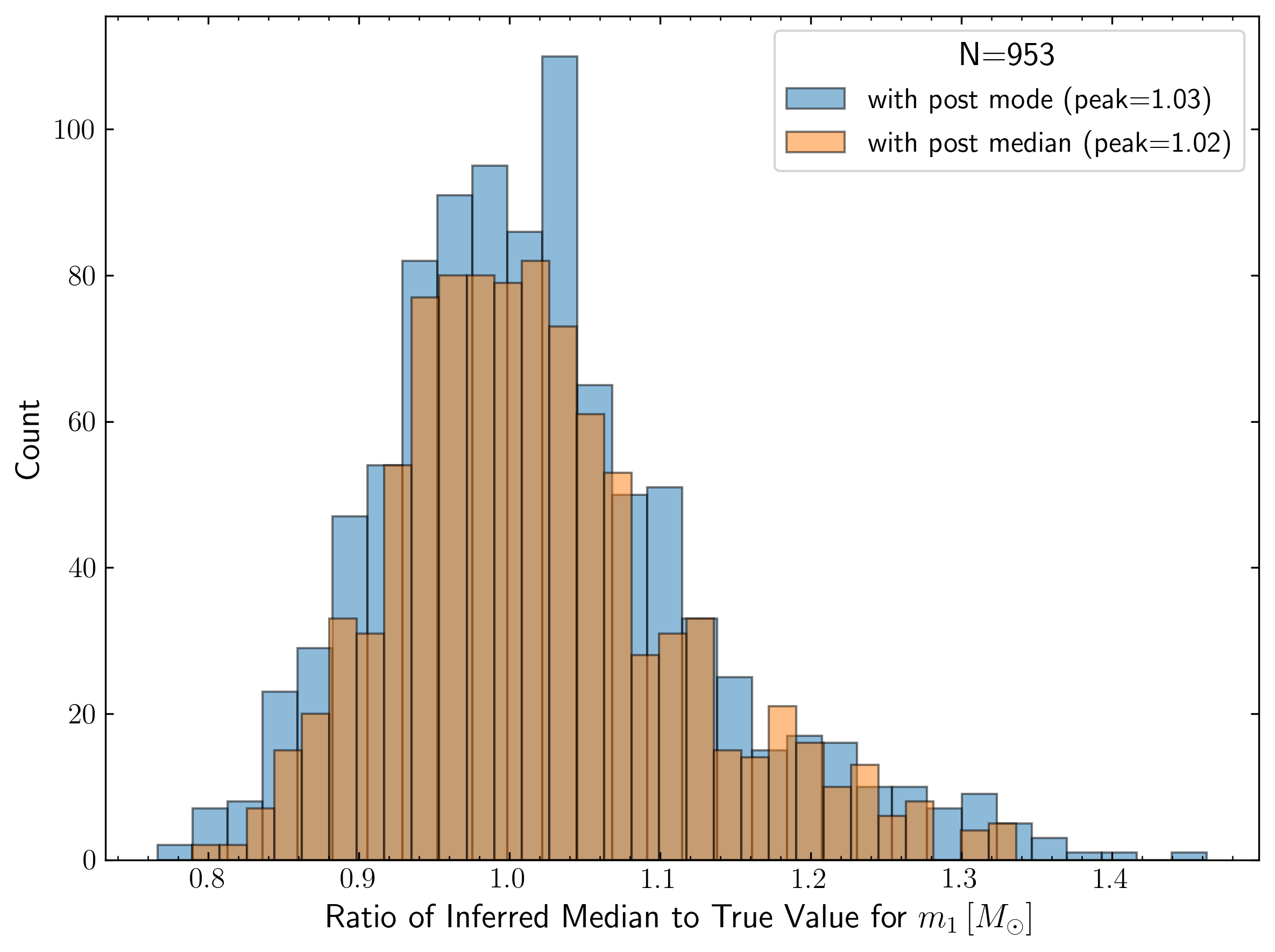}
    \includegraphics[width=\linewidth]{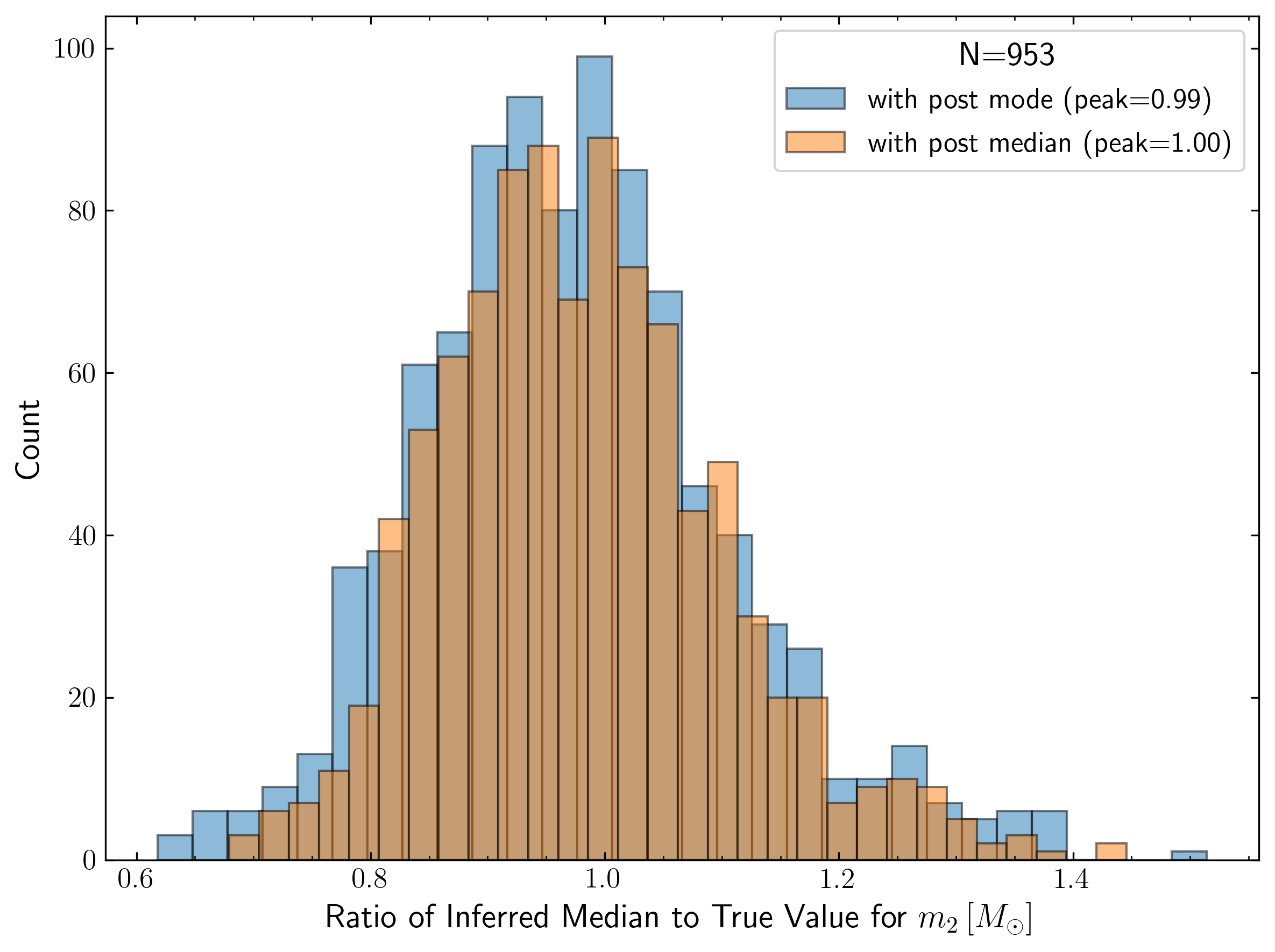}
    \caption{Distribution of the ratio of the inferred posterior mode and median value for the primary (top) and secondary (bottom) mass parameters calculated from the EOB2ML \acl{pe} runs results for injections described in \cref{fig:eob2ml-cond-pp}.}
    \label{fig:eob2ml-bias}
\end{figure}

\begin{figure}
    \centering
    \includegraphics[width=\linewidth]{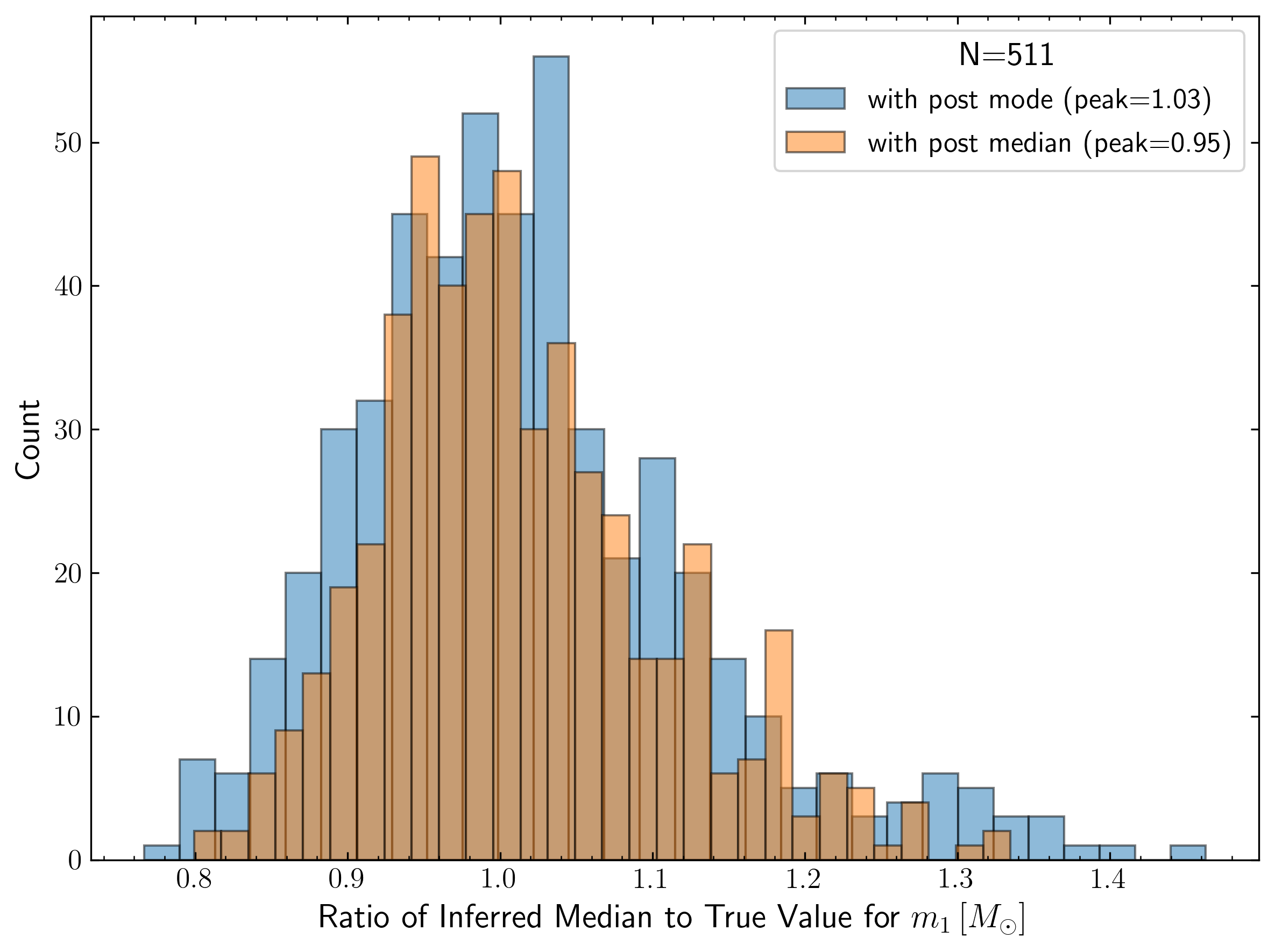}
    \includegraphics[width=\linewidth]{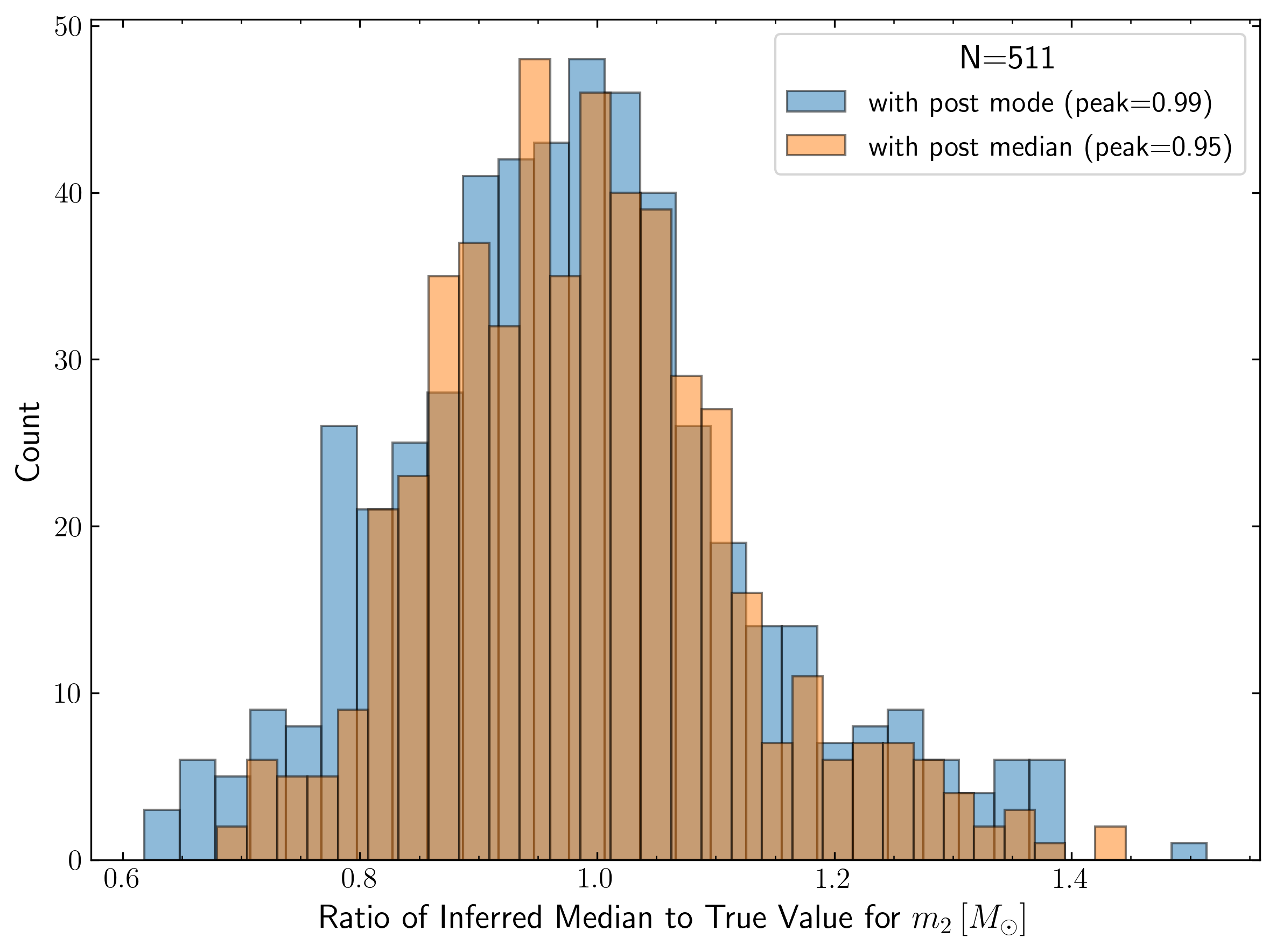}
    \caption{Distribution of the ratio of the inferred posterior mode and median value for the primary (top) and secondary (bottom) mass parameters calculated from the EOB2ML \acl{pe} runs results for injections described in \cref{fig:eob2ml-cond-d4k-pp}.}
    \label{fig:eob2ml-bias-d4k}
\end{figure}

\begin{figure}
    \centering
    \includegraphics[width=\linewidth]{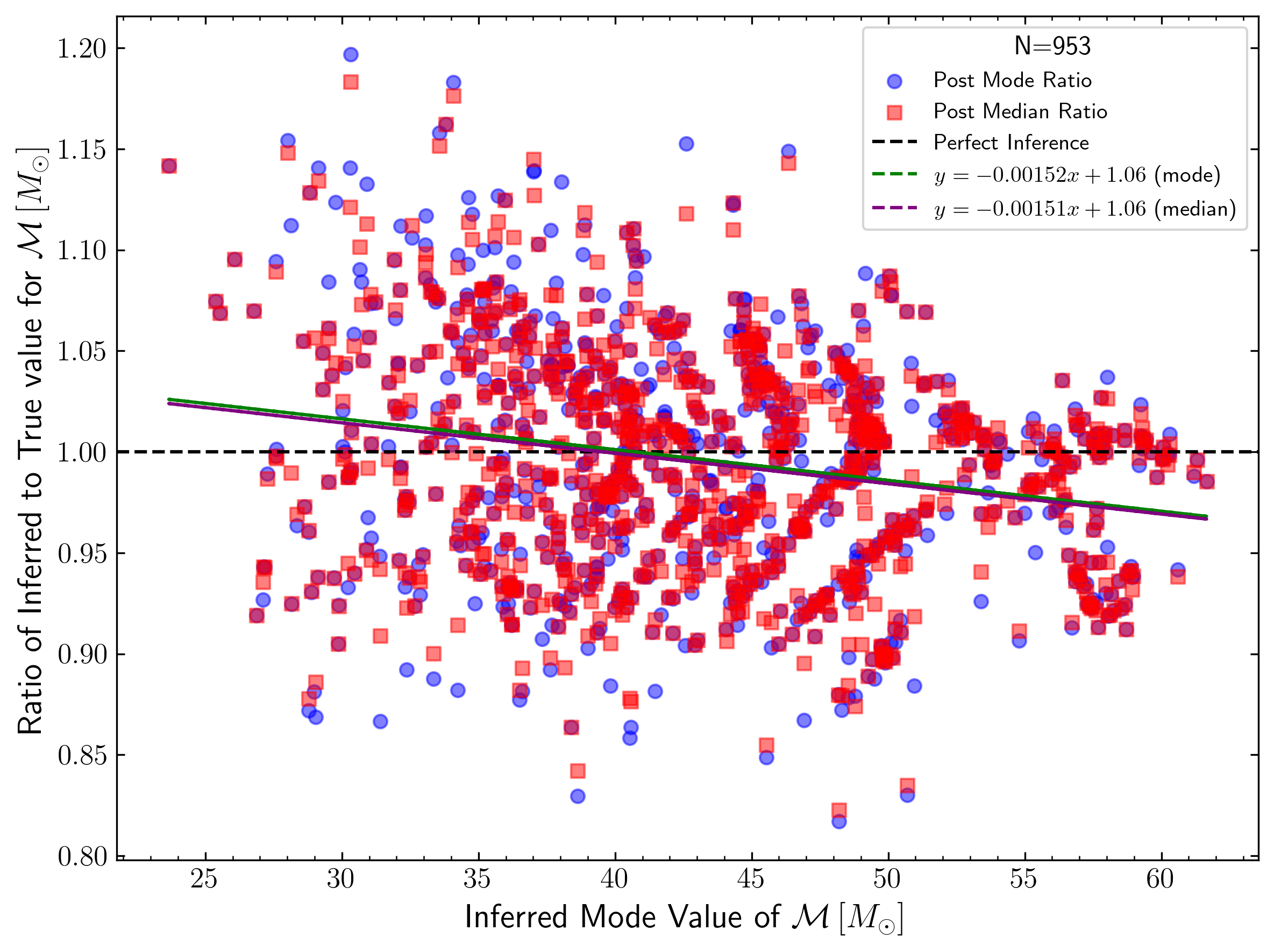}
    \includegraphics[width=\linewidth]{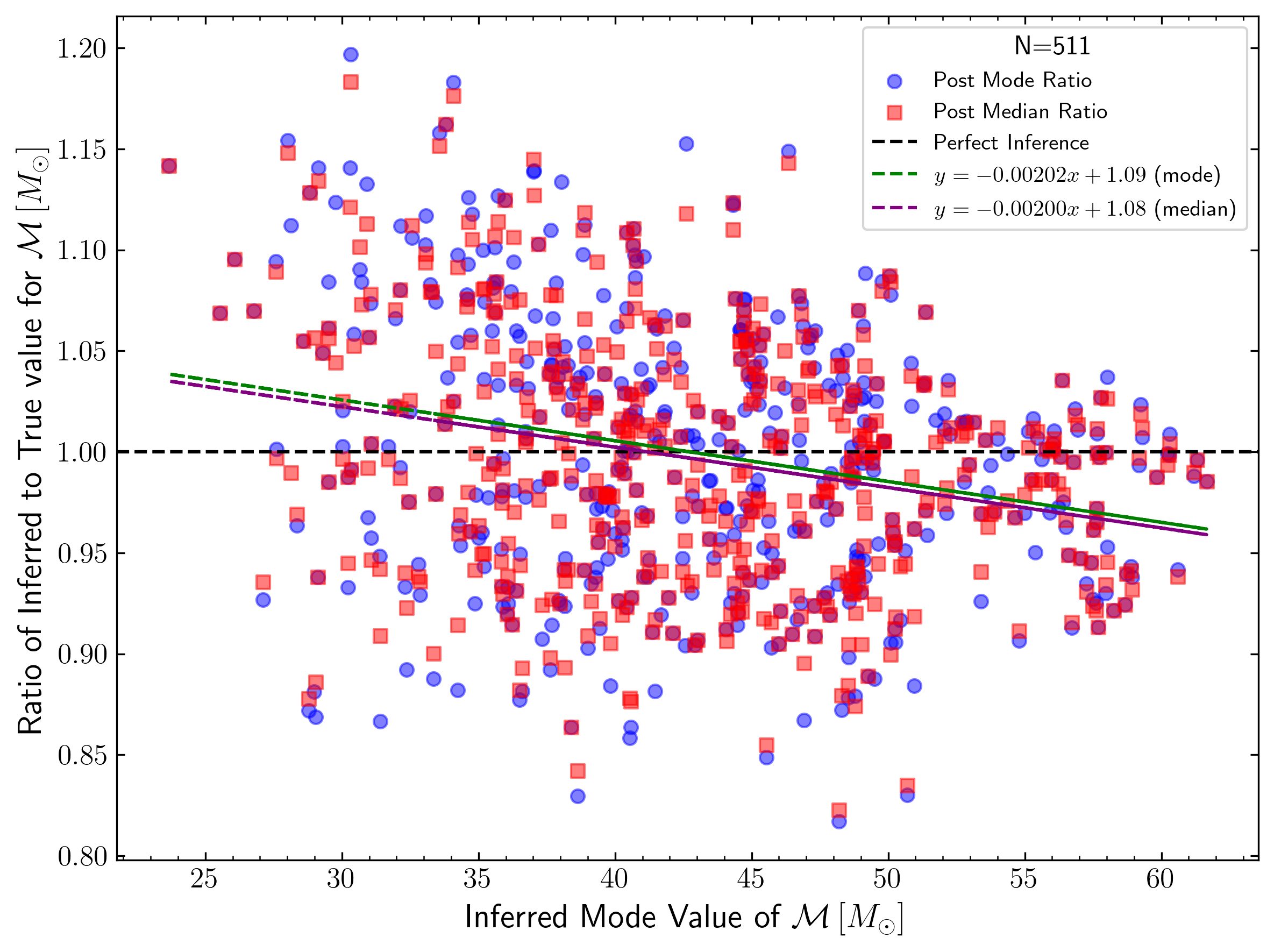}
    \caption{Inferred posterior mode versus ratio of inferred to true value for the chirp mass, calculated from the distributions shown in \cref{fig:eob2ml-bias} for $d_L=\qty{400}{\Mpc}$.
    EOB2ML \acl{pe} case (top), and in \cref{fig:eob2ml-bias-d4k} for $d_L=\qty{4000}{\Mpc}$ (bottom). The linear regression fit lines are also plotted. The ratio with the inferred posterior median is shown for completeness sake.}
    \label{fig:eob2ml-bias-chirp}
\end{figure}

For each injection $k$ and each parameter $\theta$, we define a one-dimensional posterior summary
\begin{equation}
    \hat{\theta}_{k}
    = {\rm mode}
    \left[
    P_{\rm EOB2ML}
    \left(
    \theta
    \mid d_{{\rm EOB},k}
    \right)
    \right],
\end{equation}
where the mode is computed from the one-dimensional marginalized posterior. In some comparisons we also use the posterior median,
\begin{equation}
    \tilde{\theta}_{k}
    = {\rm median}
    \left[
    P_{\rm EOB2ML}
    \left(
    \theta
    \mid d_{{\rm EOB},k}
    \right)
    \right].
\end{equation}
The posterior mode is used as a point estimate of the peak of the marginalized posterior, while the median is used as a robust central estimate. However, the full posterior distribution, rather than any single point estimate, remains the primary \acl{pe} result.

For positive-valued mass parameters, we quantify the multiplicative bias using the ratio
\begin{equation}
    R_{\theta,k}
    = \frac{
    \hat{\theta}_{k}
    }{
    \theta_{{\rm inj},k}
    },
    \qquad
    \theta \in \{m_1,m_2\}.
\end{equation}
A value $R_{\theta,k}=1$ corresponds to an unbiased marginal-mode estimate. Values above or below unity indicate a systematic over-estimation or under-estimation of the parameter. We also consider the additive residual
\begin{equation}
    \Delta \theta_{k}
    =
    \hat{\theta}_{k}
    -
    \theta_{{\rm inj},k},
\end{equation}
which is especially useful for spin parameters, because $\chi_1$ and $\chi_2$ can be negative or close to zero. 

The empirical distribution of $R_{\theta,k}$ and $\Delta\theta_k$ over the injection set provides a direct measure of the EOB2ML bias (\cref{fig:eob2ml-bias,fig:eob2ml-bias-d4k}). For example, a systematic shift of the primary-mass posterior median by approximately $10\,M_{\odot}$ relative to the injected value indicates a coherent surrogate-induced bias in the inferred mass posterior.

To test whether the EOB2ML bias is coherent and approximately correctable, we construct an empirical one-dimensional bias correction method for each parameter. For each parameter $\theta$, we can fit a linear relation between the measured bias ratio and the inferred posterior mode,
\begin{equation}
    R_{\theta,k}
    = a_{\theta}
    + b_{\theta}\hat{\theta}_{k}
    + \epsilon_{\theta,k},
\end{equation}
where $a_{\theta}$ and $b_{\theta}$ are the fitted intercept and slope, and $\epsilon_{\theta,k}$ denotes the residual scatter. The fitted bias factor evaluated at the inferred mode is therefore
\begin{equation}
    B_{\theta,k}
    = a_{\theta}
    + b_{\theta}\hat{\theta}_{k}.
\end{equation}
For each injection $k$, the one-dimensional marginalized posterior samples for parameter $\theta$ can then be corrected by dividing by this fitted bias factor,
\begin{equation}
    \theta_{k,j}^{\rm corr}
    = \frac{
    \theta_{k,j}
    }{
    B_{\theta,k}
    },
\end{equation}
where $\theta_{k,j}$ denotes the $j$-th posterior sample of parameter $\theta$ for injection $k$. Equivalently, the corrected one-dimensional marginalized posterior is
\begin{equation}
    p_{\rm corr}
    \left(
    \theta
    \mid d_{{\rm EOB},k}
    \right)
    =
    B_{\theta,k}\,
    P_{\rm EOB2ML}
    \left(
    B_{\theta,k}\theta
    \mid d_{{\rm EOB},k}
    \right),
\end{equation}
where the prefactor preserves normalization under the change of variables.

For spin parameters, when a multiplicative ratio is not stable, an additive calibration can instead be used,
\begin{equation}
    \Delta\theta_k
    =
    a_{\theta}
    +
    b_{\theta}\hat{\theta}_{k}
    +
    \epsilon_{\theta,k},
\end{equation}
with the corrected samples given by
\begin{equation}
    \theta_{k,j}^{\rm corr}
    =
    \theta_{k,j}
    -
    \left(
    a_{\theta}
    +
    b_{\theta}\hat{\theta}_{k}
    \right).
\end{equation}%
This procedure is basically a post-hoc correction of the one-dimensional marginalized EOB2ML posteriors. Thus, it does not, in any way imply that the original \ac{ml} likelihood is fully \ac{eob}-faithful. 
Rather, it tests whether the observed EOB2ML bias is coherent enough to be empirically modeled and partially removed.

\begin{figure}
    \centering
    \includegraphics[width=\linewidth]{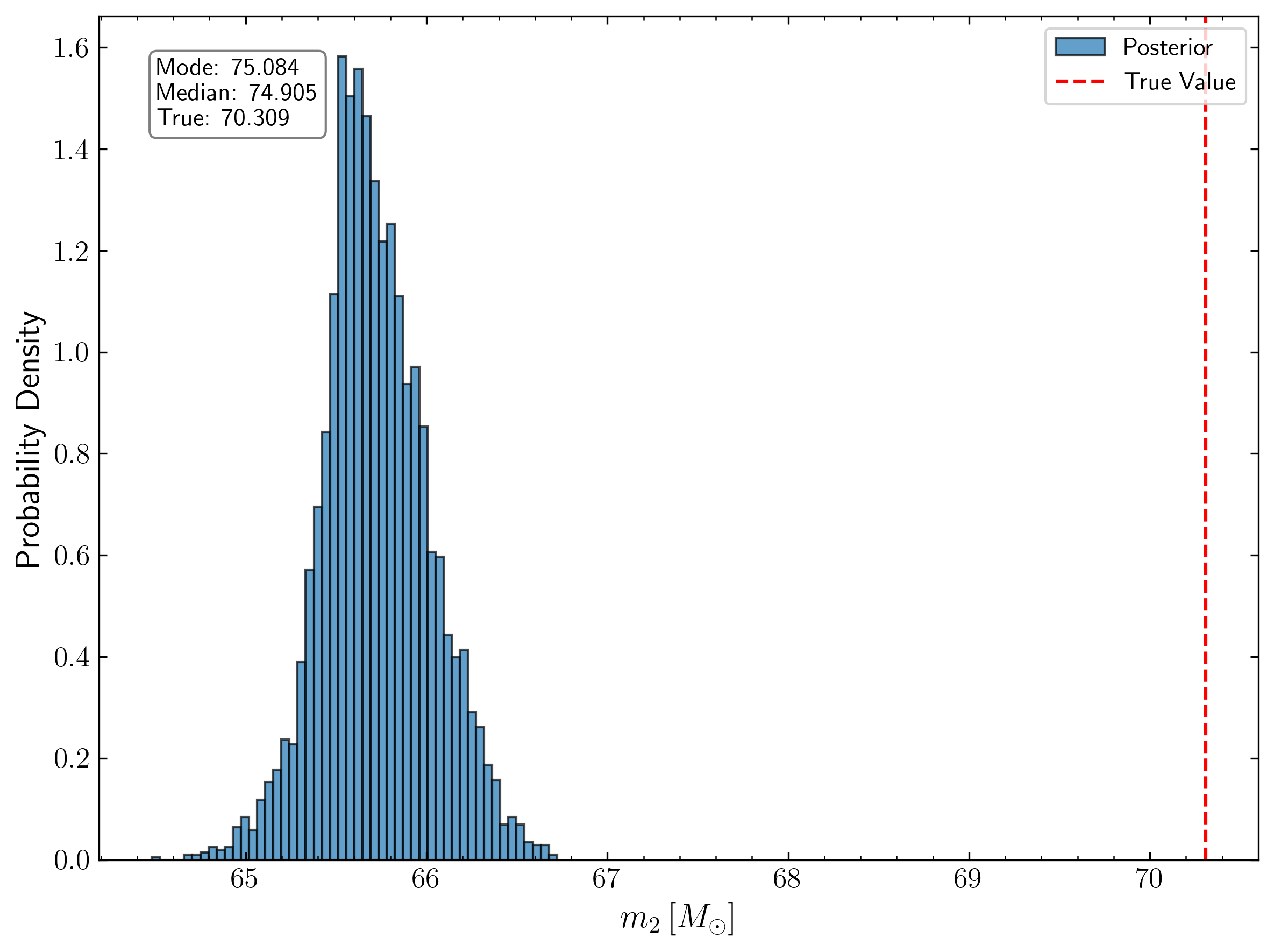}
    \includegraphics[width=\linewidth]{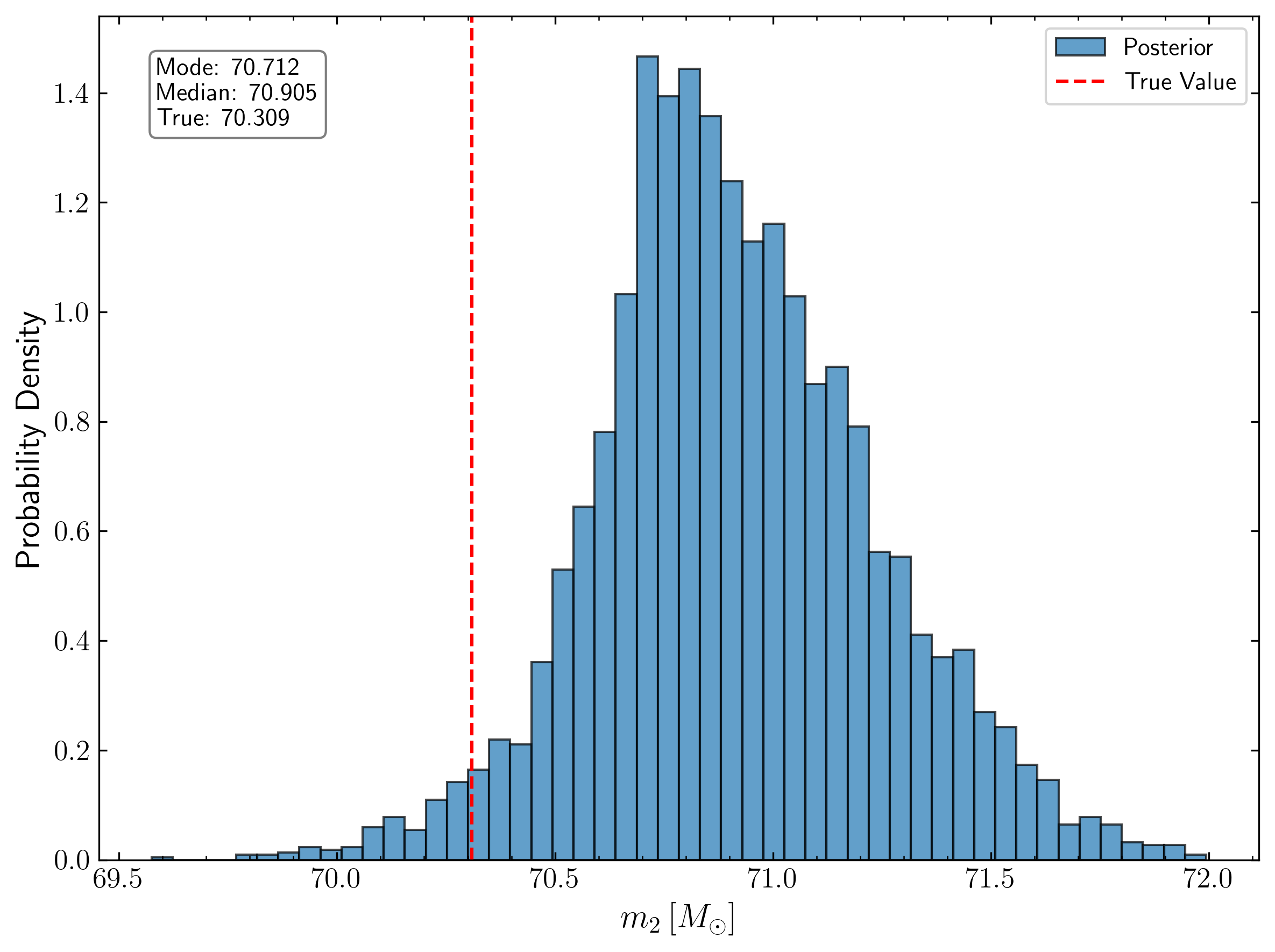}
    \caption{Biased (top) and bias-corrected (bottom) one-dimensional marginalized posteriors for the secondary mass for EOB2ML \acl{pe} run for the injection described in \cref{fig:eob2ml-cond-inj500}.}
    \label{fig:bias-correction}
\end{figure}

The derived parameter, chirp mass, is usually better constrained in \acl{pe} analysis. As \cref{fig:eob2ml-bias-chirp} shows, we find that the inferred posterior mode ratio for the chirp mass of the binary is systematically dependent, and is inversely proportional to the mass of the binary.
\cref{fig:bias-correction} shows an illustration of the bias correction procedure described in this section for the marginalized posterior for the secondary mass in the EOB2ML \acl{pe} test in \cref{fig:eob2ml-cond-inj500}. We see that the inherent bias in EOB2ML \acl{pe} is systematic and can be easily corrected for.

\subsection{Importance reweighting EOB2ML proposal to EOB2EOB target}

The EOB2ML posterior can also be interpreted as a proposal distribution for importance reweighting. The proposal posterior is
\begin{equation}
    q(\params)
    = P_{\rm ML}(\params|d_{\rm EOB})
    \propto
    \pi(\params)\like_{\rm ML}(d_{\rm EOB}|\params),
    \end{equation}
    whereas the desired target posterior is
    \begin{equation}
    p(\params)
    = P_{\rm EOB}(\params|d_{\rm EOB})
    \propto
    \pi(\params)\like_{\rm EOB}(d_{\rm EOB}|\params).
\end{equation}
Existence of an overlap in the true \ac{eob} posterior for \ac{eob} injections and the EOB2ML posterior, obtained using faster likelihood calculations with the \ac{ml} waveforms, is necessary to ensure that at least some samples lie within the desired credible region around the injection parameter.
For a posterior sample $\theta_i$ drawn from the EOB2ML proposal distribution, the corresponding importance weight can be calculated as the ratio between the target and the proposal densities.
If the priors are identical, the prior factors cancel in the importance ratio. The unnormalized importance weight assigned to a proposal sample $\theta_i\sim q(\params)$ is therefore
\begin{equation}
    w_i
    \propto
    \frac{
    \like_{\rm EOB}(d_{\rm EOB}|\theta_i)
    }{
    \like_{\rm ML}(d_{\rm EOB}|\theta_i)
    }.
\end{equation}
In practice, it is better to perform this in the log form,
\begin{equation}
    \log w_i
    = \log \like_{\rm EOB}(d_{\rm EOB}|\theta_i)
    - \log \like_{\rm ML}(d_{\rm EOB}|\theta_i).
    \label{eq:importance_weights}
\end{equation}
and with weight normalization,
\begin{equation}
    \bar{w}_i = \frac{w_i}{\sum_{j=1}^{N} w_j}.
\end{equation}
If the ML likelihood is narrowly peaked far from the true value, then the EOB2ML proposal can instead be generated from a tempered likelihood as,
\begin{equation}
    q_{\alpha}(\params)
    \propto
    \pi(\params)\like_{\rm ML}(d_{\rm EOB}|\params)^{\alpha},
    \qquad 0<\alpha\leq 1,
\end{equation}
and then the correct importance weights become
\begin{equation}
    \log w_i
    = \log \like_{\rm EOB}(d_{\rm EOB}|\theta_i)
    - \alpha
    \log \like_{\rm ML}(d_{\rm EOB}|\theta_i).
    \label{eq:tempered_importance_weights}
\end{equation}
The role of the tempering parameter is to broaden the proposal posterior. Smaller values of $\alpha$ correspond to a higher effective temperature and can increase overlap with the \ac{eob} target posterior.

The \ac{eob} posterior expectation of a function $f(\params)$ is then approximated by
\begin{equation}
    \mathbb{E}_{\rm EOB}[f(\params)]
    \simeq \sum_{i=1}^{N}
    \bar{w}_i f(\theta_i).
\end{equation}
The quality of the weighted sample is quantified by the effective sample size
\begin{equation}
    N_{\rm eff} = \frac{1}{\sum_{i=1}^{N}\bar{w}_i^2}.
    \label{eq:neff}
\end{equation}
If all samples have equal weights, then $N_{\rm eff}=N$. If a single sample dominates the weight, then $N_{\rm eff}\approx 1$. Therefore, $N_{\rm eff}/N$ is a direct diagnostic of proposal-target overlap. A small effective sample fraction will indicate that the \ac{ml} proposal does not sufficiently cover the \ac{eob} target posterior, and an equal-weight reweighted posterior obtained by rejection or resampling will be unreliable.
We applied the importance-reweighting method for some of the posteriors from our EOB2ML \acl{pe} run, however, we find that the sufficient posterior overlap only occurs at low \acp{snr}. In the next section, we discuss this behaviour in more detail.


\section{Discussion on Parameter Estimation with injection at different SNR}

\subsection{Expected one-dimensional likelihood comparison as a function of SNR}

A useful diagnostic is to compare the one-dimensional likelihood profiles in a single parameter, for example $m_1$, while fixing over the remaining parameters. Let
\begin{equation}
    \ell_{\rm EOB}(m_1)
    = \log \like_{\rm EOB}(d_{\rm EOB}|m_1,\hat{\params}_{\setminus m_1})
\end{equation}
and
\begin{equation}
    \ell_{\rm ML}(m_1)
    = \log \like_{\rm ML}(d_{\rm EOB}|m_1,\hat{\params}_{\setminus m_1}),
\end{equation}
where $\hat{\params}_{\setminus m_1}$ denotes either fixed injected values or locally optimized values of the remaining parameters. In the ideal case, both likelihoods peak at the same value of $m_1$. If the \ac{ml} waveform is biased relative to the \ac{eob} waveform, the \ac{ml} likelihood peak is shifted away from the \ac{eob} peak.

At high \ac{snr}, the likelihood is narrow, and even a small separation between the \ac{eob} and \ac{ml} likelihood maxima can produce negligible posterior overlap. At lower \ac{snr}, the likelihood broadens, and the proposal posterior obtained from the EOB2ML run may overlap more strongly with the true EOB2EOB posterior. This improved overlap makes importance reweighting more likely to produce a finite and useful effective sample size.

This behavior can be understood using a simple Gaussian approximation. Suppose that the \ac{eob} target and \ac{ml} proposal in one dimension are approximately
\begin{align}
    p(m_1)
    &= \normal(m_1;\mu_{\rm EOB},\sigma^2),
    \nonumber \\
    q(m_1)
    &= \normal(m_1;\mu_{\rm ML},\sigma^2),
\end{align}
with a mean displacement $\Delta m_1 = \mu_{\rm ML}-\mu_{\rm EOB}$. Since statistical uncertainty typically scales as $\sigma(\snr)\propto \snr^{-1}$,
the separation in units of posterior width grows as
\begin{equation}
    \frac{|\Delta m_1|}{\sigma(\snr)}
    \propto
    \snr |\Delta m_1|.
\end{equation}
Thus, even if the absolute parameter bias is approximately fixed, its significance increases with \ac{snr}. In this simplified case, the effective sample fraction for importance weighting decreases approximately exponentially with the squared separation:
\begin{equation}
    \frac{N_{\rm eff}}{N}
    \sim
    \exp\left[
    - \frac{
    \Delta m_1^2
    }{
    \sigma(\snr)^2
    }
    \right].
\label{eq:neff_gaussian_scaling}
\end{equation}
It should be noted though that this is not an exact prediction for the full multidimensional \acl{gw} posterior, but it captures the expected trend: proposal-target overlap is largest at low \ac{snr} and rapidly decreases as the signal becomes louder.

The \ac{snr} value at which importance reweighting becomes useful depends on the waveform mismatch and on the size of the waveform-induced parameter bias. A rough waveform-accuracy criterion is given by \cref{eq:mismatch_snr_condition}. For our current surrogate waveform model that has $\mismatch\sim10^{-2}$, one expects unbiased or strongly overlapping EOB2ML proposals only at relatively low \ac{snr}, roughly $\snr\lesssim10$. For a more accurate surrogate with $\mismatch\sim10^{-4}$, the same heuristic suggests that useful overlap may persist up to $\snr\sim100$. However, since the actual behavior depends on the local structure of the likelihood and on correlations among masses and spins, these values are only approximate.


\subsection{Cost-based metric for the usefulness of the EOB2ML proposal}

A practical motivation for EOB2ML sampling followed by \ac{eob} reweighting is to reduce the number of expensive \ac{eob} likelihood evaluations. We can define a EOB-likelihood cost factor as
\begin{equation}
    \mathcal{R}_{\rm cost}
    = \frac{
    N_{\rm ML}^{\rm rw}
    }{
    N_{\rm EOB}^{\rm direct}
    },
    \label{eq:cost_ratio}
\end{equation}
where $N_{\rm EOB}^{\rm rw}$ is the number of \ac{eob} likelihood evaluations required to reweight samples from the \ac{ml} proposal, and $N_{\rm EOB}^{\rm direct}$ is the number of \ac{eob} likelihood evaluations required for a direct EOB2EOB \acl{pe} run. Values $\mathcal{R}_{\rm cost}<1$ indicate that the reweighting approach is cheaper in terms of \ac{eob} likelihood calls. The corresponding likelihood speedup is
\begin{equation}
    \mathcal{S}_{\rm ML}
    = \frac{1}{\mathcal{R}_{\rm cost}}
    = \frac{
    N_{\rm EOB}^{\rm direct}
    }{
    N_{\rm ML}^{\rm rw}
    }.
\end{equation}

However, a low cost factor is meaningful only if the reweighted posterior is accurate. We should therefore impose a usefulness criterion based on the effective sample size and posterior agreement:
\begin{equation}
    N_{\rm eff}
    \geq
    N_{\rm eff}^{\rm min},
    \qquad
    D\left(
    P_{\rm rw}(\params),
    P_{\rm EOB}(\params)
    \right)
    \leq
    \epsilon,
    \label{eq:usefulness_criterion}
\end{equation}
where $D$ is a posterior-distance measure, such as a Kolmogorov-Smirnov distance in one-dimensional marginals, or a difference in credible intervals. If these conditions are not satisfied, the reweighting result should not be considered useful, regardless of the nominal cost factor.

Let the effective sample fraction be $f_{\rm eff} = N_{\rm eff} / N$. Then the number of proposal samples needed to achieve a target effective sample size is approximately, $N_{\rm ML}^{\rm rw} \simeq N_{\rm eff}^{\rm min} / f_{\rm eff}$, and therefore,
\begin{equation}
    \mathcal{R}_{\rm cost}
    \simeq
    \frac{
    N_{\rm eff}^{\rm min}
    }{
    f_{\rm eff} N_{\rm EOB}^{\rm direct}
    }.
    \label{eq:cost_ratio_feffective}
\end{equation}
This equation shows explicitly that the efficiency of the method is controlled by the proposal-target overlap. At low \ac{snr}, where the \ac{eob} and \ac{ml} posteriors are broad and overlapping, $f_{\rm eff}$ may be large and the advantage factor can be much smaller than unity. At high \ac{snr}, $f_{\rm eff}$ can become extremely small, causing the reweighting cost to approach or exceed the cost of a direct EOB2EOB run.

\begin{figure}
    \centering
    \includegraphics[width=\linewidth]{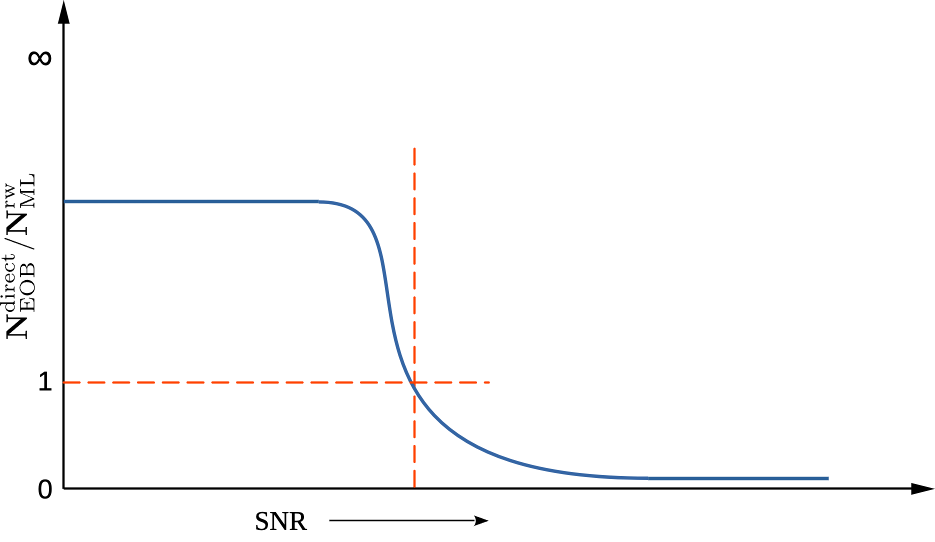}
    \caption{Schematic sketch showing the expected trend in likelihood speedup gain, given by the ratio of number of likelihood evaluations required in an EOB2EOB \acl{pe} run over the number of likelihood evaluations required when reweighting samples from the EOB2ML proposal distribution.
	At low \ac{snr}, the EOB2ML proposal posterior is broad and may overlap significantly with the EOB2EOB target posterior, enabling efficient importance reweighting. At high \ac{snr}, waveform-systematic bias becomes more significant relative to the statistical uncertainty, the effective sample fraction decreases, and the cost of likelihood evaluation increases. The reweighting strategy is useful only when it satisfies both a likelihood speedup condition, $\mathcal{S}_{\rm ML}>1$, and an accuracy condition, such as $N_{\rm eff}\geq N_{\rm eff}^{\rm min}$.}
    \label{fig:p2-snr-nlike}
\end{figure}

\cref{fig:p2-snr-nlike} shows the expected qualitative behavior of the proposal usefulness as a function of \ac{snr}. At low \ac{snr}, the posterior width is large and even a biased \ac{ml} likelihood can retain overlap with the \ac{eob} target posterior. As \ac{snr} increases, the posterior narrows approximately as $\snr^{-1}$, while waveform-systematic error shifts become increasingly significant in units of the width of the posterior. Therefore, the effective sample fraction decreases, resulting in a decrease in the likelihood speedup gained.

\subsection{Expected improvement from a more accurate waveform model}

The results above suggest a clear route for improvement. The EOB2ML posterior bias arises because the current \ac{ml} waveform model is not accurate enough to reproduce the \ac{eob} waveform family at the level required for unbiased parameter estimation. Improving the waveform accuracy should reduce the separation between the \ac{eob} and \ac{ml} likelihood maxima, $\left| \theta_{\rm ML}^{\star} - \theta_{\rm EOB}^{\star} \right|$,
and should increase the overlap between the \ac{ml} proposal posterior and the \ac{eob} target posterior. This, in turn, should increase $N_{\rm eff}/N$ and reduce the EOB-likelihood advantage factor in \cref{eq:cost_ratio_feffective}.

The mismatch--\ac{snr} criterion in \cref{eq:mismatch_snr_condition} provides a useful target. As we have seen, if a \acl{ml} waveform model has typical mismatch $\mismatch\sim10^{-2}$, it is expected to show waveform-induced posterior bias for moderate and high-\ac{snr} signals. A model with $\mismatch\sim10^{-4}$ would be expected to remain useful up to substantially higher \acp{snr}. Therefore, improving the \acl{ml} waveform model accuracy by one to two orders of magnitude can qualitatively change the behavior of EOB2ML parameter estimation and make importance reweighting a practical correction strategy.

This motivates the development of a higher-accuracy surrogate or a residual-corrected waveform model. In such a model, the surrogate waveform can be written schematically as
\begin{equation}
h_{\rm corr}(\params) = h_{\rm ML}(\params) + \Delta h_{\rm res}(\params),
\end{equation}
where $\Delta h_{\rm res}$ is a learned correction to the residual between the baseline \ac{ml} waveform and the \ac{eob} waveform. If the corrected waveform reduces the mismatch sufficiently, then the true EOB2ML proposal posterior should become closer to the EOB2EOB posterior, the importance weights should become less concentrated, and the effective sample size should increase.


\section{Summary}

In this work, we developed a \ac{cae} waveform generation model for \texttt{SEOBNRv4} aligned-spin gravitational waveforms, following along the lines of \citet{garg2026a}. However, in contrast to their model, our model is fully-deterministic and learns a fixed latent-space vector for each set of input source parameters and the associated waveform. This model is trained to predict the amplitude and phase series as a function of time. These outputs of the waveform generation are calibrated with a second-stage residual calibration model, bringing the final waveform polarization mismatch to $\order{10^{-2}}$.

We then introduced a waveform conditioning step, which tapers the start and end of the reconstructed waveforms, and embeds these \qty{1}{\second} long signals into \qty{8}{\second} long data segment. This conditioning step is necessary to ensure that no frequency-domain edge effects affect the frequency bins of the data once the time-domain data segment is converted to frequency-domain via Fourier transformation in downstream \acl{pe} tasks. Finally, with these conditioned-waveforms we perform extensive \acl{pe} tests, quantify the \acl{pe} bias for the case when \ac{ml} waveforms are used to infer properties of \ac{eob} target waveforms, and propose strategies to correct this bias.

The conditioned-waveform ML2ML \acl{pe} analysis produces a diagonal \ac{pp}-plot, showing that the inference machinery is internally calibrated when the injection and recovery waveform families are identical. On the other hand, the EOB2ML study addresses the following question: how posteriors change when data generated by an \ac{eob} waveform model are recovered using the \ac{ml} waveform model. The conditioned-waveform EOB2ML analysis shows biased posteriors and a non-diagonal \ac{pp}-plot, indicating that the conditioned \ac{ml} surrogate does not exactly reproduce the conditioned \ac{eob} likelihood surface. We quantify this posterior bias using the distribution of ratios or residuals between inferred posterior summaries and injected parameter values. 

We find that a simple empirical correction based on a linear fit to the bias ratio as a function of the inferred posterior mode can partially correct the one-dimensional marginalized posteriors.
This post-hoc correction is useful as a diagnostic because it shows that the EOB2ML bias has a coherent structure. However, it is not a replacement for improving the waveform model. A fully reliable EOB2ML likelihood surrogate should reduce the bias at the waveform level, such that the raw EOB2ML posteriors are calibrated without requiring an empirical posterior correction.

We also discussed importance reweighting as a possible correction strategy for cases where ML2ML \acl{pe} runs give diagonal \ac{pp}-plots, but the EOB2ML runs do not. In this approach, samples from the \ac{ml} posterior are reweighted using the ratio between the \ac{eob} and \ac{ml} likelihoods. The success of this strategy depends on the overlap between the \ac{ml} proposal posterior and the \ac{eob} target posterior. This overlap can be quantified by the effective sample size $N_{\rm eff}$. If the EOB2ML posterior is too far from the EOB2EOB target posterior, the importance weights collapse and the reweighted posterior is unreliable. Conversely, at low enough \ac{snr}, or for sufficiently accurate \ac{ml} waveforms, the two posteriors may overlap and reweighting can recover the \ac{eob} posterior using fewer \ac{eob} likelihood evaluations than a direct EOB2EOB run.
A future production-ready implementation of our methods can aim to reweight the EOB2ML results, allowing further interpretation of posterior overlaps, waveform-systematic biases, and proposal efficiencies.

Finally, we introduced a cost-based metric for evaluating the usefulness of EOB2ML proposals. The relevant quantity is the ratio between the number of \ac{eob} likelihood evaluations required for reweighting over the number required for a direct EOB2EOB analysis. This metric must be considered together with an accuracy criterion, such as a minimum effective sample size or agreement with EOB2EOB posterior marginals. The expected behavior is that the reweighting approach is most useful at low \ac{snr}, where the posteriors are broad and overlapping, and becomes less efficient at high \ac{snr}, where waveform-systematic biases are resolved by the likelihood.

Thus, in summary, this work establishes both the feasibility of integrating \acl{ml} surrogate waveforms into the \texttt{Bilby} \acl{pe} framework and the need for careful posterior-level validation before using such models as production-level waveform surrogates or proposal distributions for full \acl{pe} runs.
The ML2ML results with conditioned waveform outputs show that fast \ac{ml} waveforms can be incorporated into a Bayesian \acl{pe} pipeline. 
The EOB2ML results show that bias correction remains a requirement before the surrogate waveforms can be used as an unbiased replacement for \ac{eob} target waveforms, especially at high-\ac{snr} data.

This work further motivates the development of a higher-accuracy or residual-corrected models in future. If the accuracy of reconstructed waveforms itself can be increased, say by incorporating a third fine-tuning stage, the downstream \acl{pe} analysis will get better automatically. This will be one of the aims of our follow-up works. However, in this work we have shown that the parameter inference bias arising due to slightly less-accurate waveform model is coherent and can be properly estimated and corrected for. Still, our present work focuses on relatively simple dominant mode \texttt{SEOBNRv4} waveforms, and creating \acl{ml} surrogate waveform models for more complicated waveforms, for e.g. those from eccentric binary sources, can another of our next targets.

\begin{acknowledgments}
We would like to thank Feng-Li Lin, Yuta Michimura, Hideyuki Tagoshi, Masaki Ando, Naoki Yoshida, Michiko Fuji, and Takahiro S. Yamamoto for useful discussions.
S.G. is grateful to Alvin Li and the Croucher foundation for their supporting in funding visits to Hong Kong, during which a part of this work was performed.
S.G. would also like to thank Kavli-IPMU for the CD3-Google seed fund grant, and Soichiro Morasaki and Kipp Cannon for funding through the JSPS KAKENHI research grant number 23H04893.
We acknowledge computational resources made available at the LDG-CIT cluster in Caltech and at the RESCEU-BBC cluster in UTokyo.
\end{acknowledgments}


\bibliographystyle{apsrev4-2}
\bibliography{phd-paper-II.bib}

\end{document}